\documentclass{aa}  
\usepackage{graphicx}
\usepackage{subcaption}
\usepackage{txfonts}
\usepackage{xcolor}
\usepackage{hyperref}
\usepackage{booktabs}
\usepackage[utf8]{inputenc}  
\usepackage[T1]{fontenc}     
\usepackage{textcomp}        
\usepackage{soul}
\usepackage{cancel}
\usepackage{capt-of}
\newcounter{letteredsub}[subsubsection]
\renewcommand{\theletteredsub}{\thesubsubsection.\Alph{letteredsub}}

\newcommand{\letteredsubsection}[1]{
  \refstepcounter{letteredsub}
  \paragraph*{\theletteredsub\quad #1}
  \addcontentsline{toc}{paragraph}{\protect\numberline{\theletteredsub}#1}
\mbox{}\\  
\noindent\hspace*{-3pt}
}

\newcommand{\vsini}{$v\sin{i}$}
\newcommand{\vmic}{$v_{\rm mic}$}
\newcommand{\vmac}{$v_{\rm mac}$}
\newcommand{\vr}{$v_{\rm rad}$}
\newcommand{\kms}{km\,s$^{-1}$}
\newcommand{\rholeo}{$\rho$\,Leo}

\begin{document}

   \title{Photometric and spectroscopic variability of the blue supergiant {\rholeo}}

   \author{V.A. Checha \inst{1}\thanks{E-mail: vitalii.checha@ut.ee}\
          \and
          A. Aret\inst{1}
          \and 
          I. Kolka\inst{1}
          \and
          T. Liimets\inst{1}
          \and
          I. Araya\inst{2}
          \and
          A. Christen\inst{3}
          \and
          G. F. Avila Marín\inst{3}
          \and
          R. S. Levenhagen\inst{5}
          \and
          L.~Cidale\inst{6,7}
          \and
          T. Eenmäe\inst{1}
          \and
          G. Hajiyeva\inst{4,5}
          \and
          Ü. Kivila\inst{1}
          \and
          V. Mitrokhina\inst{1}
          \and
          H. Ramler\inst{1}
          \and
          T. Verro\inst{1}
          }

\institute{Tartu Observatory, University of Tartu, Observatooriumi 1, T\~{o}ravere, 61602, Estonia.
    \and
    Centro Multidisciplinario de F\'isica, Vicerrector\'ia de Investigaci\'on, Universidad Mayor, 8580745 Santiago, Chile.
    \and
    Instituto de Estadística, Universidad de Valparaiso, Blanco 951, Valparaíso, Chile.
    \and
    Division of Physics of stellar atmospheres and magnetism, Shamakhy Astrophysical Observatory, Shamakhy, 5600 Azerbaijan.
    \and
    Graduate School of Science, Art and Technology, Khazar University, 41 Mahsati Str., AZ 1096, Baku, Azerbaijan.
    \and
    Departamento de Física, Universidade Federal de São Paulo, Rua Prof. Artur Riedel, 275, 09972-270, Diadema, SP, Brazil.
    \and
    Instituto de Astrofísica La Plata, CCT La Plata, CONICET-UNLP, Paseo del Bosque S/N, B1900FWA La Plata, Argentina.
    \and
    Departamento de Espectroscopía, Facultad de Ciencias Astron\'omicas y Geofísicas, Universidad Nacional de La Plata (UNLP), Paseo del Bosque S/N, B1900FWA La Plata, Argentina.
\\}

   \date{Received 20~June~2025; accepted 14~December~2025}

\abstract
   {The post-main-sequence evolution of massive stars remains poorly understood, particularly in the case of blue supergiants. As key drivers of the dynamical and chemical evolution of galaxies, massive stars warrant detailed investigation during this complex evolutionary stage. Hot supergiants exhibit pronounced photometric and spectroscopic variability, typically in the form of quasi-periodic rather than strictly periodic variations.}
   {We investigated the variability patterns of the evolved B-type star {\rholeo} to determine its properties, identify the underlying physical processes, and constrain its evolutionary stage. We   combined extensive long-term datasets of spectroscopic and photometric observations from various sources. These include data from the TESS and Kepler space telescopes, as well as observations from the 1.5~m telescope in Estonia.} 
   {We analysed the data using the generalized Lomb–Scargle periodogram, the Lomb–Scargle periodogram with pre-whitening, and the weighted wavelet Z-transform. To determine the fundamental parameters of {\rholeo}, we fitted synthetic line profiles computed with the \textsc{FastWind} code to the HARPS spectrum. We used the \textsc{zpektr} code to infer the stellar rotation inclination angle.}  
   {The \ion{He}{I}\, 6678.151\,\AA\ line profile exhibits significant changes in radial velocity and, consequently, in its moment values. We identify a set of periods and harmonics ranging from $\sim 0.8$ to $\sim 35$ days. Some periods remain nearly constant, while others appear and disappear from one observing season to another. A comparison of spectroscopic and photometric data, along with the shape of the phase curves, helps to constrain the nature of several periods. In particular, the $\sim$11-day period is attributed to stellar rotation, while the $\sim$17-day period is associated with radial pulsations.}  
   {Despite their quasi-periodic nature, most periods are observable across multiple observing seasons.  Based on the fairly wide range of detected periods, {\rholeo} is likely on the blue loop of its evolution, following the red supergiant stage.}

\maketitle

\section{Introduction}\label{S-int}

Over the past few decades, asteroseismology has become a powerful tool for probing stellar interiors. Studying the variability of blue supergiants (BSGs) is particularly important as their pulsational behaviour provides direct clues about the internal structure and the physical processes that shape the evolution of massive stars (see the review by \citealt{2023Ap&SS.368..107B}). Investigating objects such as {\rholeo} therefore contributes to the broader effort to link observed surface variability with the underlying physics of stellar interiors.

{\rholeo} (\object{HD 91316}) is a slowly rotating BSG of spectral class B1\,Iab \citep{1997yCat..72840265H}. It has a mass of approximately 22\,$M_{\odot}$, a radius of 32\,$R_{\odot}$, an effective temperature of $T_{\rm eff}=20\,260$\,K, and a projected rotational velocity of \vsini\,$=49$\,\kms \citep{2014A&A...562A.135S}. {\rholeo} is classified as an $\alpha$\,Cygni-type variable, exhibiting quasi-periodic photometric and spectroscopic variability on timescales of days (e.g. \citealt{1988ApJS...68..319L}, \citealt{refId1}). The binary nature of this star has been extensively discussed in the literature (e.g. \citealt{1976A&A....48..245D}, \citealt{2024AJ....168...28T}  and references  therein); however, no conclusive evidence has been found. 

The stellar wind parameters of {\rholeo}, including mass loss and wind variability, have been studied by \cite{2006A&A...446..279C}. In addition, it is known to display non-radial pulsations \citep{2016MNRAS.458.1604K}. The connection between the pulsation pattern and the evolutionary stage of BSGs has been established by \citet{2013MNRAS.433.1246S}, and yet detailed information on the variability of {\rholeo}, the existing periods, and a comprehensive comparison of these periods is still lacking. 

Long-term spectroscopic observations of hot stars such as {\rholeo} reveal diverse structures within their atmospheres, with lifetimes ranging from fractions of an hour to several days. For OB stars, variations in line profiles are often regular or quasi-regular, with long-term periodic variations spanning several days to tens of days. These periodic changes are frequently associated with variations in mass loss, reflecting the dynamic nature of stellar atmospheres (\citealt{1980S&T....60R.418C, 1989nos..book.....U, 2010aste.book.....A}). Longer periods are more likely to be associated with radial pulsation modes. To investigate this further, \cite{2013MNRAS.433.1246S} extended their calculations of pulsation instabilities; they modelled pulsation patterns that follow the evolution of massive stars from the main sequence to the red supergiant (RSG) stage and beyond. They conducted these calculations for stars with initial masses up to 25 $M_{\odot}$, considering both non-rotating stars and stars with initial rotation rates of 40\% of their critical rotation velocity.

As described by \cite{2021WSAAA..12..234K}, studying stellar pulsations using photometry presents two main challenges, both related to the stellar wind and its variability over time. The stellar wind influences the observed brightness of the star, distorting the light curve. As a result, photometry alone cannot be used to study pulsations and must be combined with spectroscopic time series.

A detailed variability analysis of {\rholeo} was carried out by \citet{10.1093/mnras/sty308}, who combined 80 days of high-cadence K2 halo photometry with 1800 days of high-resolution HERMES spectroscopy. All detected variability was confined to frequencies below 1.5 d$^{-1}$. The dominant frequency, $f_{\rm rot}=0.0373$ d$^{-1}$ (period \textasciitilde26.8 days), was attributed to rotational modulation at the base of an aspherical stellar wind and interpreted as the stellar rotation signature.   In addition, the combined photometric and spectroscopic data revealed low-amplitude photospheric velocity variations in the range \textasciitilde0.2–0.6 d$^{-1}$, which were attributed to large-scale travelling gravity waves in the stellar envelope.

The study of variability in the line profiles of BSGs is one of the most effective methods for exploring the structure of their atmospheres. These stars exhibit multi-periodic photometric and spectroscopic variations on timescales ranging from about 1 to 100 days. These variations are typically attributed to high‑order non‑radial pulsations (NRPs), specifically g‑modes. A review of the available literature on line profiles in the spectra of early‑type stars has shown that reliable detection of both irregular (stochastic) and regular (periodic, often related to NRP) profile changes requires achieving a signal‑to‑noise ratio (S/N) of 300 or higher, particularly in the region of the lines under study \citep{2007ARep...51..920K}. In addition, a high time resolution, of the order of $\sim15$ minutes, is necessary to Nyquist‑sample the fastest g‑modes, to trace the hour‑scale bumps as they sweep across the line profile, and still reach S/N$\gtrsim$300 per exposure without smearing out these rapid variations.

The strength of our study lies in the combination of K2 and TESS photometry with two seasons of high-cadence spectroscopic observations obtained in close temporal proximity. Such near-simultaneous coverage enables a consistent investigation of the photometric and spectroscopic variability of blue supergiants, where pulsation modes are short-lived, frequencies drift with time, and the observed signals are quasi-periodic.

For this study we used the weighted wavelet Z-transform (WWZ) \citep{1996AJ....112.1709F} and  Lomb–Scargle analysis methods (\citealt{1976Ap&SS..39..447L,1982ApJ...263..835S,2009A&A...496..577Z}) to identify frequencies in the light curves from the Kepler/K2\footnote{\url{https://science.nasa.gov/mission/kepler/}} and TESS\footnote{\url{https://tess.mit.edu/}} space telescopes, and in our spectroscopic observations. The 2017 spectra were obtained just before the K2 campaign, while in 2022 one spectrum preceded, two coincided with, and the rest closely followed the TESS observations.

In addition to analysing the temporal variability, we employed detailed computational modelling to constrain the inclination of the stellar rotation axis, a parameter that had not been determined for this star in previous studies. Together, these approaches provide new insights into the complex variability of {\rholeo} and help to distinguish stable periodic signals from those that evolve with time.

\section{Observations and data processing}\label{S-obs}

\subsection{TESS photometry}\label{S-tessobs}
    TESS observed {\rholeo} from 6~November to 30~December~2021, in sectors 45 and 46, in the short-cadence mode (2 min exposures), for a total of almost 7000 measurements covering 52.44 days. We extracted the light curve of {\rholeo} from the TESS full-frame images downloaded from the MAST\footnote{\url{https://archive.stsci.edu/missions-and-data/tess}} archive.
    Due to the target's saturating magnitude, {\rholeo} flux was distributed over a relatively large number of TESS pixels. Therefore, the corresponding aperture for the stellar signal and background was carefully selected according to the guidelines provided by \cite{2021AJ....162..170H} and \cite{2021ApJS..257...53L}.
    Our custom pixel mask (aperture) for {\rholeo} in TESS sector 45 is shown in Fig.~\ref{TESS_aperture} in Appendix~\ref{timeseries}. Within this aperture (radius $\sim 2'$), there are eight additional faint stars (G\,=\,15.4--20.3 mag, Gaia DR3) that have no measurable effect on the  {\rholeo} light curve. The stellar signal was measured within the aperture marked by pure-white pixels with counts exceeding approximately four times the background level.
    
    We cannot rule out the possibility that the target is also variable on timescales longer than the duration of a single sector ($\sim 26$ days). Hence, to better match the instrumental time series of the two sectors, we also extracted the light curve of the comparison star 48\,Leo, which is positioned inside the same CCD field as {\rholeo}. Consequently, we present the photometric TESS time series of {\rholeo} (Fig.~\ref{TESS_lc}) as differential magnitudes against the stable comparison star 48\,Leo, which has a flux scatter of less than $\sim 1$~ppt. 
    
One of the most striking features of the TESS light curve is its distinct sawtooth shape. Interestingly, the same pattern has been observed in K2 photometry \citep[][Fig. 1]{10.1093/mnras/sty308}, further supporting its validity.

\subsection{Kepler K2 photometry}\label{S-K2obs}

{\rholeo} was observed with the Kepler space telescope in the framework of the K2 Campaign 14 (K2) from 30~May to 19~August~2017, just after our spectroscopic observations in  Season 2017 (Sect.~\ref{S-specobs}). 
The K2 dataset consists of long-cadence data (29.4 min) covering 79.64~d. 
For our analysis, we used the K2 halo photometric light curve of {\rholeo} published by \citet{10.1093/mnras/sty308}. The light curve was digitized from their Figure 1, which presents the brightness variations extracted using the first photometric mask (upper left panel in their Figure A1). This is the same aperture configuration that \citet{10.1093/mnras/sty308} employed for their frequency analysis, thereby ensuring that our study is based on an identical dataset and allowing a direct comparison of the results.

\subsection{Spectroscopy}\label{S-specobs}

Archival photometric data were complemented with the spectroscopic observations obtained with the Tartu Observatory's (TO) 1.5 m telescope AZT-12. The long-slit spectrograph ASP-32 in the Cassegrain focus was used with a 1800 line mm\textsuperscript{-1} diffraction grating providing spectra in the wavelength range from 6300\,\AA{} to 6730\,\AA{} with a S/N $\sim300-400$ and resolution $R \approx 10\,500$. In total, data were collected during 150 observing nights between 2014 and 2024, resulting in approximately 2700 spectra. 
  
A typical sequence of observations covers a duration of 2--4.5 hours during the night with a typical exposure time of approximately 300 seconds (ranging from 150 to 800 seconds depending on weather conditions). For this article we focused on the two most important observational seasons: winter--spring 2017 (referred to as Season 2017) and autumn 2021--spring 2022 (referred to as Season 2022). The data for season 2017 were collected just before the K2 photometry, between 4~January and 16~May~2017, comprising a total of 369 spectra over 20 nights. Spectra for Season 2022 were obtained between 1~November~2021 and 15~May~2022, totalling 1106 spectra over 39 nights, collected mostly shortly after the TESS photometry. We note that one spectrum was obtained before the TESS photometry, on 1~November~2021, and two spectra during TESS observations, on 21~November~2021 and 9~December~2021. The data were reduced using the \textsc{IRAF}\footnote{IRAF was written at the National  Optical Astronomy Observatory, which was operated by the Association of Universities for Research in Astronomy (AURA) under cooperative agreement with the National Science Foundation.}  \footnote{\url{https://iraf-community.github.io/}} software (\citealt{1986SPIE..627..733T,1993ASPC...52..173T}), employing standard routines from the \textsc{noao}, \textsc{imred}, and \textsc{ccdred} packages.
First, bias and flat-field corrections were applied to the spectra;  dark correction was not necessary because of adequate cooling of the CCD. 
Next, wavelength calibration was performed using the ThAr lamp frames obtained before and after the target exposures. 
To further check and refine the stability of the wavelength scale during the observing run, telluric lines near 6570 \AA\ were used. Additionally, heliocentric correction was applied to the spectra. Continuum normalization was achieved by fitting a first-order (cubic spline) function. 
A summary of the observing periods and the number of spectroscopic data points for all datasets used in this work is given in Table~\ref{tab:observations}, while the full log of observations is presented in Table~\ref{tab:Spectra_all2} in Appendix~\ref{S-appendixobs}. 

\begin{table}[!ht]
        \centering
        \caption{Spectroscopic and photometric observing periods analysed in this work in chronological order.
        }
        \begin{tabular}{l c c c c}
            \hline\hline
            Season     & Period of observations & Data points \\
            \hline
            2017  & 2017-01-04 - 2017-05-16 &  369  \\
            K2    & 2017-05-30 - 2017-08-19 &  1879 \\

            2022  & 2021-11-01 - 2022-05-15 &  1106 \\
            TESS  & 2021-11-06 - 2021-12-30 &  6862 \\
            \hline
        \end{tabular}
    \label{tab:observations}    
\end{table}

In addition to our own long-term spectroscopic observations, the pipeline-reduced high-resolution spectrum, R = 115 000, obtained with the High Accuracy Radial Velocity Planet Searcher (HARPS) attached to the 3.6 m telescope at La Silla Observatory, observed on 12~February~2006 (HJD-2453778.618849) was downloaded from the ESO Science Archive Facility.\footnote{Based on data obtained from the ESO Science Archive Facility with DOI \url{https://doi.eso.org/10.18727/archive/33} under the ESO programme 60.A-9036} The spectrum has a \hbox{S/N $\sim300$} and covers a wavelength range of $3781-6913$\,\AA{}. 

\section{Results}\label{S-res}

\subsection{Stellar parameters of \rholeo}\label{S-modelstpar}
To reassess the stellar parameters of \rholeo, the high-resolution HARPS spectrum observed on 12~February~2006 was used. This particular spectrum was selected because the spectral lines of interest appeared symmetric, indicating a period of weak stellar-wind activity. Choosing a spectrum obtained under such calm wind conditions minimizes the influence of wind-related distortions on the line profiles, and thus ensures the most reliable determination of the atmospheric parameters.

First, we derived the parameters responsible for the broadening of the spectral lines; the macroturbulent velocity, \vmac; and the projected rotational velocity, $v{\sin i}$. For this purpose, we analysed the silicon line, \ion{Si}{III} $\lambda$4552, using the {\sc iacob-broad} tool,\footnote{\url{https://research.iac.es/proyecto/iacob/pages/en/useful-tools.php}} developed within the framework of the IACOB project \citep{2014A&A...562A.135S}. The chosen diagnostic line is well isolated, exhibits a symmetric photospheric profile, and shows minimal wind interference, making it particularly suitable for this analysis.  

The {\sc iacob-broad} tool combines a Fourier transform and a goodness-of-fit analysis to disentangle rotational and macroturbulent broadening, and adopts a radial–tangential macroturbulent profile with equal radial and tangential components, following \citet{2014A&A...562A.135S}. The code provides two independent estimates of the projected rotational velocity, $v{\sin i}$(FT) and $v{\sin i}$(GOF), whose agreement serves as an internal reliability check. The tool also provides uncertainties from the GOF analysis based on the corresponding $\chi^2$ distributions, and these values are reported in \hbox{Table \ref{tab:Phys.param}}.

\begin{figure}
\centering
\includegraphics[width=\linewidth]{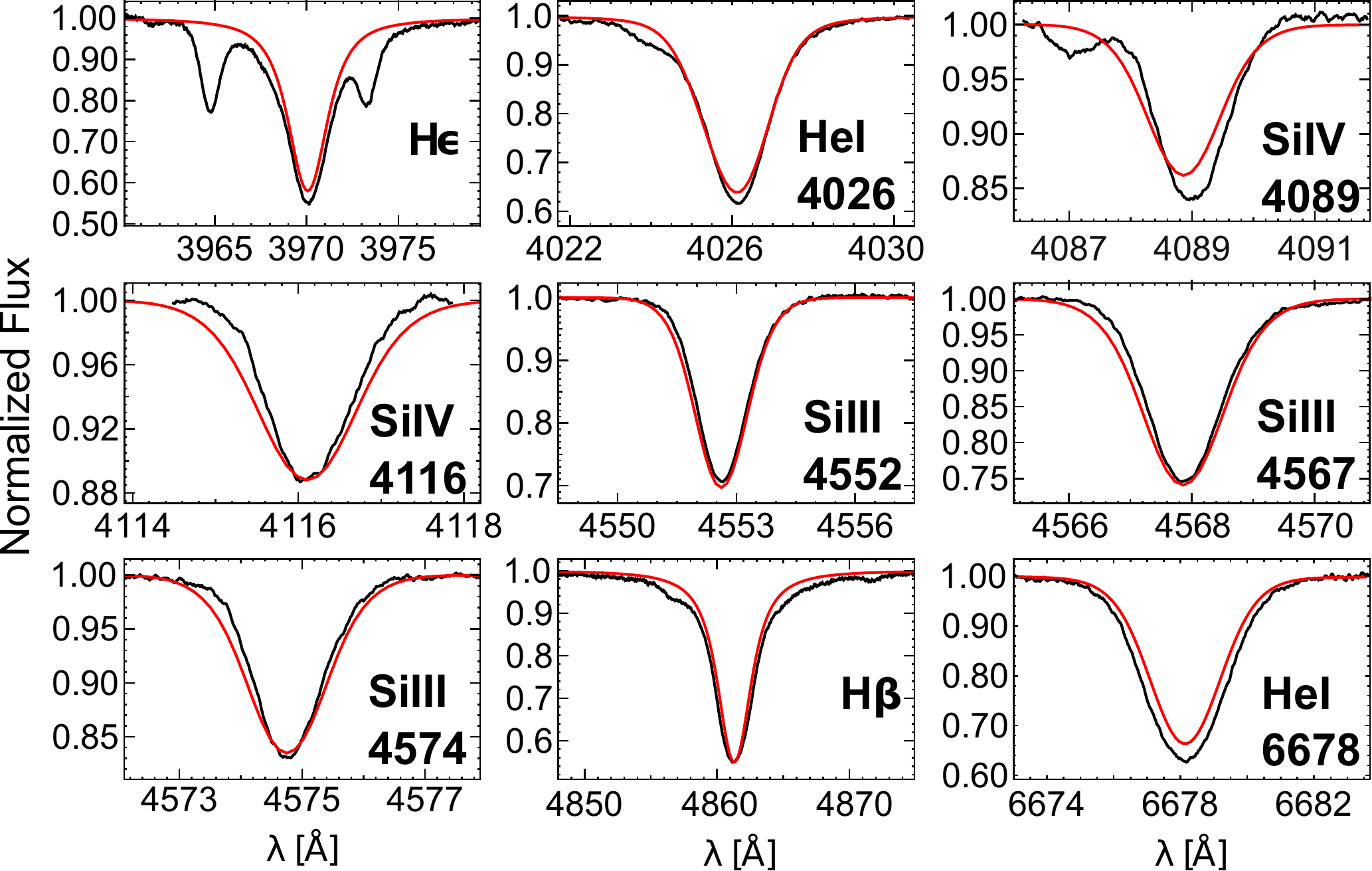}
\caption{Comparison between the observed spectrum (black line) and the best-fitting \textsc{FastWind} model (red line).}
    \label{fig:FastWind1}
\end{figure}

Next, we used a set of spectral lines to determine the rest of the stellar parameters with the \textsc{FastWind} v10.6 code \citep{Puls}. Uncertainties in $T_{\rm eff}$, $\log g$, $M$ [$M_{\odot}$], and \vmic were estimated by varying the derived parameters with \textsc{FastWind}. The fit to the observed spectrum is shown in Fig.~\ref{fig:FastWind1}. The results of our computations and measurements, along with previously established parameters, are presented in \hbox{Table \ref{tab:Phys.param}}.

\begin{table}[ht]
    \centering
    \caption{Physical parameters of \rholeo.}
    \begin{tabular}{c c c }
        \hline\hline
        Parameter           & Value & Reference   \\
        \hline
        $T_{\rm eff}$ [kK]       & $23\pm1$ &  This work  \\
                            &   20.26 - 22 &     [1,2] \\
        $M$ [$M_{\odot}$]      & 22   & [1]  \\
        $R$ [$R_{\odot}$]      & 32 & [1]  \\
        $\log L$ [$L_{\odot}$]      & 5.18 - 5.47 & [1,2]  \\
        $\log g$                & $2.6\pm0.1$ &  This work  \\
                            &   2.55        & [2] \\
        $\dot{M}$ [$M_{\odot}$yr$^{-1}$]           & 3.5E-7 & This work \\
                            & 6.3E-7 & [1] \\
        $v_{\infty}$ [\kms]  & 1\,110 & [2]  \\
        $\beta$             & 1.0  & [2]  \\
        $v{\sin i}$ [\kms]        & $49\pm2.4$ & This work  \\
                            & 75    & [1,2] \\
        \vmac\ [\kms]          & $76\pm2.4$ & This work  \\
        \vmic\ [\kms]         & $20\pm5$   & This work  \\
                            &   10   & [2] \\
        \vr\ [\kms]          & 42 & This work  \\
                            &   42  & [1] \\
        V [mag]              & 3.85 & [2]  \\
        \hline\\[-8pt]
        \multicolumn{3}{l}{[1] \cite{10.1111/j.1365-2966.2004.07799.x}}\\
        \multicolumn{3}{l}{[2] \cite{2006A&A...446..279C}}\\
    \end{tabular}
    \label{tab:Phys.param}
\end{table}

\subsection{Time series analysis}

Due to uneven sampling and gaps in the analysed time series, the following widely used techniques were employed in this study to perform the frequency analysis: generalized Lomb–Scargle (GLS) periodogram, generalized Lomb–Scargle periodogram with pre-whitening (GLSp), and weighted wavelet Z-transform (WWZ).

\underline{GLS periodogram}  \citep{2009A&A...496..577Z}. This approach is  an extension of the classical Lomb–Scargle method \citep{1976Ap&SS..39..447L,1982ApJ...263..835S}; it accounts for measurement uncertainties and allows for a floating mean. In GLS the underlying model is a full linear least-squares fit of a sinusoid plus offset, which provides an unbiased treatment of the data errors. As shown by \citet{2009A&A...496..577Z}, this approach increases the statistical power of the period search and yields more accurate and robust frequency estimates, especially for uneven sampling and noisy time series. For each dataset, the most significant frequencies were identified from the GLS power spectrum and visually inspected to confirm that they are not artefacts of the time sampling. The significance was assessed via the Lomb–Scargle false-alarm probability (FAP), using a 1\% global FAP threshold derived from both the analytic Baluev approximation and a bootstrap with 5000 resamples. In the GLS periodograms, peaks above the FAP threshold are considered significant.

\underline{GLSp}. This approach refines the standard Lomb–Scargle analysis by iteratively removing significant frequencies, thus enabling the identification of weaker signals.

\underline{WWZ} \citep{1996AJ....112.1709F}. In addition to the GLS periodogram, we also applied the time-frequency WWZ method to study the temporal evolution of the detected frequencies. The method is a widely used for analysing periodic signals in time series data \citep[e.g.][]{2023Galax..11...69A,2024A&A...689A..35C,2025MNRAS.tmp.1744W}. This technique allows the identification of transient or time-dependent periodicities by computing localized power spectra across sliding time windows.

\subsubsection{Analysis of the K2 light curve}\label{S-anaphotK2}

\begin{figure}
    \centering
    \includegraphics[width=\linewidth]{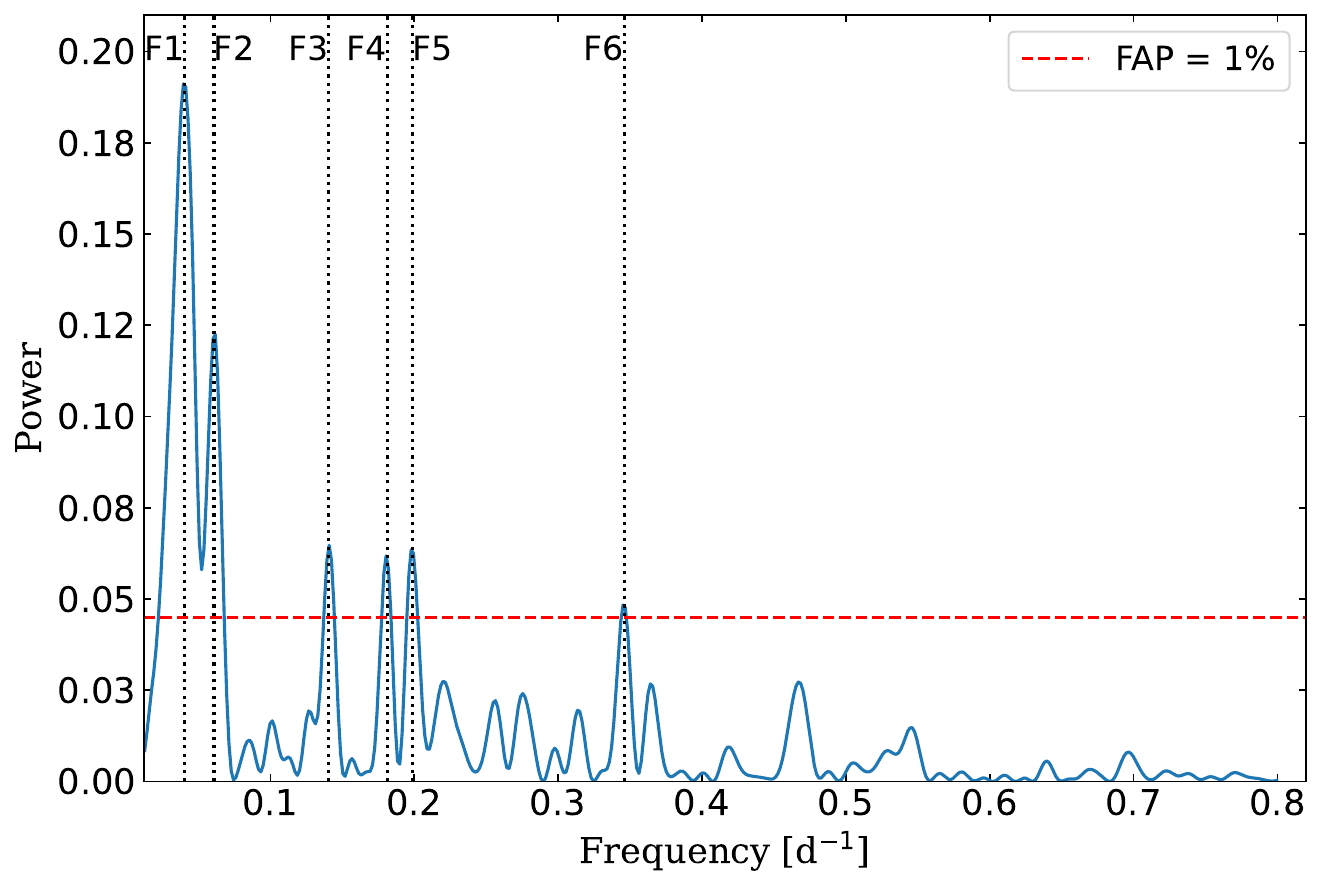}
    \caption{Lomb–Scargle power spectrum of the K2 photometry. The first six most powerful frequencies are shown. The dominant frequency (0.0405 d$^{-1}$) refers to a period of 24.7 days.} 
\label{fig: LoSc K2}
\end{figure}

As mentioned in Sect.~\ref{S-K2obs}, the light curve for K2 was digitized from \citet{10.1093/mnras/sty308}. However, we   applied different frequency analysis methods. The results are presented in Table~\ref{tab:K2_combined_dec_period}.
First, we implemented a GLS for the K2 photometry. The length of the K2 photometry was 78.64 days, and hence we can take a range of periods for analysis of up to 39.32 days. The power spectrum is presented in Fig.~\ref{fig: LoSc K2}. We note that the vast majority of frequencies lie in the long-wave region (periods over 1.25 days). 

The clear predominance of low frequencies visible in the Lomb–Scargle spectrum is of physical origin and correlates with the stochastic low-frequency (SLF) variability intrinsic to massive stars (\citealt{2019NatAs...3..760B,2020A&A...640A..36B,2023Ap&SS.368..107B}). SLF variability in the photometry of massive stars is currently believed to be inherent to the stars themselves, arising from internal gravitational waves and turbulent convection. 

\setlength{\tabcolsep}{5pt}
\begin{table*}[h!]
    \captionsetup{justification=raggedright,singlelinecheck=false}
    \centering
    \caption{Periods detected in the K2 light curve.}
    \label{tab:K2_combined_dec_period}
    \begin{tabular}{lccccccccccc}
        \toprule
        \multicolumn{4}{c}{GLS} &
        \multicolumn{4}{c}{GLSp} &
        \multicolumn{4}{c}{WWZ} \\
        \cmidrule(lr){1-4} \cmidrule(lr){5-8} \cmidrule(lr){9-12}
        N & Frequency & Period & Ident. &
        N & Frequency & Period & Ident. &
        N & Frequency & Period & Ident. \\
        & [d$^{-1}$] & [d] & & & [d$^{-1}$] & [d] & & & [d$^{-1}$] & [d] & \\
        \midrule
        $F_1$ & 0.0405 [1] & $24.7 \pm 0.6$  & --               &
        $F_1$ & 0.0402 [1] & $24.85 \pm 0.61$ & --               &
        $F_1$ & 0.0395 [1] & $25.3 \pm 5.4$  & --             \\

        $F_2$ & 0.0608 [2] & $\mathbf{16.4} \pm 0.2$  & --               &
        $F_2$ & 0.0521 [3] & $19.21 \pm 0.20$ & --              &
        $F_2$ & 0.0610 [2] & $\mathbf{16.4} \pm 1.8$  & --             \\

        $F_3$ & 0.1406 [3] & $\mathbf{7.1} \pm 0.4$   & --               &
        $F_3$ & 0.0622 [2] & $\mathbf{16.07} \pm 0.25$ & --              &
        $F_3$ & 0.1000 [5] & $\mathbf{10.0} \pm 1.0$  & --    \\

        $F_4$ & 0.1812 [5] & $5.5 \pm 0.2$   & $F_1 + F_3$      &
        $F_4$ & 0.1410 [4] & $\mathbf{7.10} \pm 0.40$  & --     &
        $F_4$ & 0.1408 [4] & $\mathbf{7.1} \pm 0.5$   & $F_1 + F_3$    \\

        $F_5$ & 0.1989 [4] & $5.0 \pm 0.2$   & $F_2 + F_3$      &
        $F_5$ & 0.1980 [5] & $5.05 \pm 0.20$  & $F_2 + F_4$     &
        $F_5$ & 0.1923 [3] & $5.2 \pm 0.5$   & $F_2 + F_4$      \\

        $F_6$ & 0.2205 [7] & $4.5 \pm 0.0^*$ & --               &
        $F_6$ & 0.3462 [6] & $2.90 \pm 0.00 $ & $F_4 + F_5$  &
        --    & --     & --              & --             \\

        $F_7$ & 0.3459 [6] & $2.9 \pm 0.0^*$ & $F_3 + F_4$      &
        --    & --     & --      & --      &
        --    & --     & --              & --             \\
        \bottomrule
    \end{tabular}
    \tablefoot{Results of the analysis of three methods: GLS, GLSp, and WWZ, arranged by decreasing period in each column. The number in brackets [N] denotes the rank of the period by detected power. An asterisk (*) indicates an uncertainty of less than $0.04$~days. In this and the following tables, persistent periods detected in all datasets are highlighted in \textbf{bold}.}
\end{table*}

\begin{figure}
    \centering
    \includegraphics[width=\linewidth]{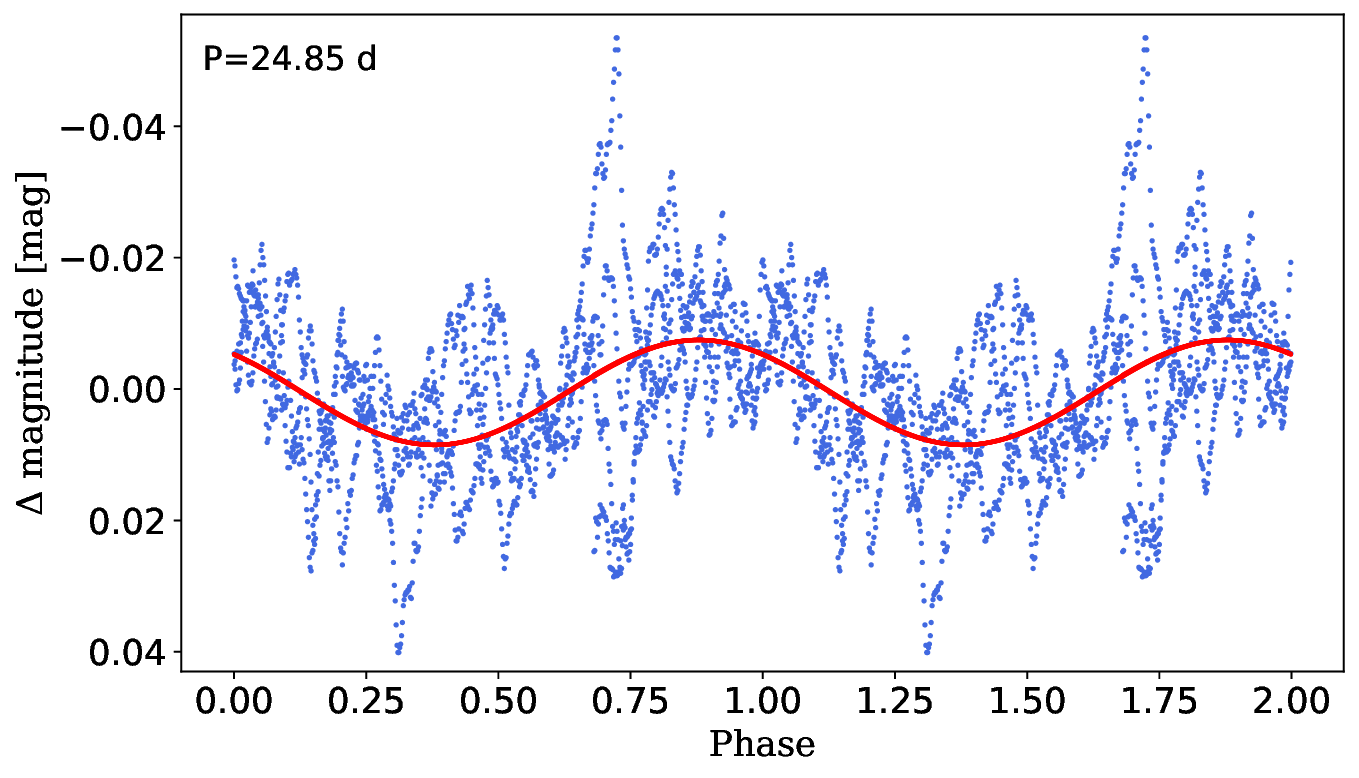}
    \caption{Phase diagram of K2 photometry with a period of 24.85 days.} 
    \label{fig:K2 25d}
\end{figure}

\begin{figure*}[h]
    \centering
    \includegraphics[width=\linewidth]{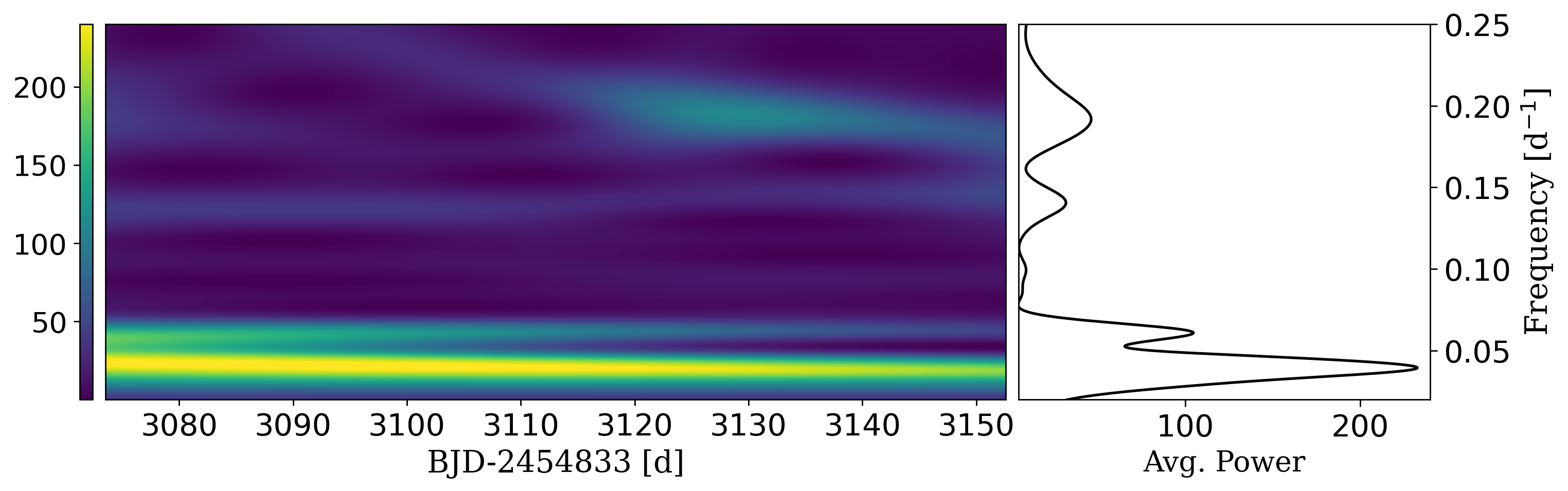}
    \caption{WWZ scalogram for K2 photometry. The wavelet power is indicated by the colour (see colour bar at left). The lower bright stripe corresponds to the period of $\sim$25 days. The black solid line in the right panel shows the time-averaged wavelet power.} 
    \label{fig:K2 wwz}
\end{figure*}

To verify the periods obtained and find the exact value of the periods, the GLSp method was applied to the data.
In Table \ref{tab:K2_combined_dec_period} we present a set of frequencies and the sub-harmonics that we found. We note that frequency $F_5 \sim F_2 + F_4$ and frequency $F_6 \sim F_4 + F_5$.
Figure \ref{fig:K2 25d} shows the phase diagram for the most significant period, 24.85 $\pm$ 0.61 days, which is consistent within the uncertainties with the \mbox{26.8-day} period reported by \citet{10.1093/mnras/sty308}.  The second-strongest peak corresponds to a period of 16.07 ± 0.25 days (see discussion in Sect. \ref{discus}).

In addition, we performed a WWZ analysis and obtained a scalogram (Fig.~\ref{fig:K2 wwz}) to study the change in frequency and amplitude over time. The 25.3-day period is clearly visible as the brightest stripe at the bottom of the WWZ scalogram. This period is not stable; it increases towards the end of the observations. We  also note that the power of this period is more pronounced at the beginning. The 10-day period also exists, but it is much weaker. The shorter periods appear as spots.

\subsubsection{Analysis of the TESS light curve}\label{S-anaphotTESS}

\begin{figure}
    \centering
    \includegraphics[width=\linewidth]{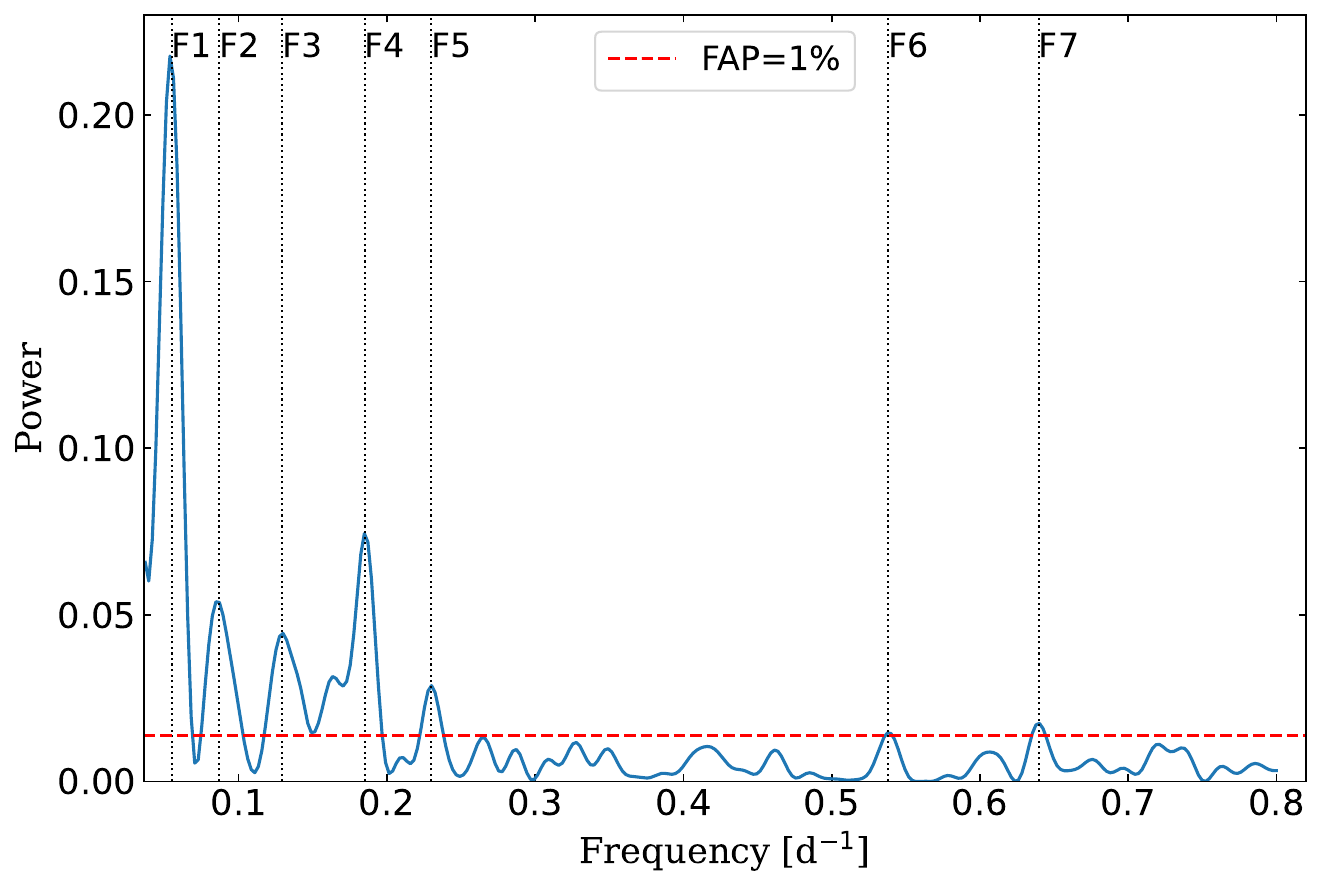}
\caption{Lomb–Scargle power spectrum of the TESS photometry. The first seven most powerful frequencies are shown. The dominant frequency of 0.055 d$^{-1}$ refers to a period of 18.2 days.}
\label{fig:TESS_GLS}
\end{figure}

The first results of our analysis are shown in the GLS frequency scalogram presented in Fig.~\ref{fig:TESS_GLS}. This figure offers a detailed view of the dominant frequencies present in the light curve and also includes the FAP level. 
We identify the set of significant frequencies, up to 0.63 d$^{-1}$; they are listed in Table~\ref{tab:TESS_combined_sorted}. 
We note that the evident SLF pattern is similar to the K2 photometry (Fig.~\ref{fig: LoSc K2}). The dominant frequency is around 0.055~d$^{-1}$, or the period of $P \sim18.2$ days. However, not all significant frequencies are independent. Many signals are sub-harmonics or combinations of the independent frequencies; we note that 
$F_4 \sim F_1 + F_3$ and $F_5 \sim 4 \times F_1$.

\begin{table*}[h!]
    \captionsetup{justification=raggedright,singlelinecheck=false}
    \centering
    \caption{Periods detected in TESS photometry.}
    \label{tab:TESS_combined_sorted}
    \begin{tabular}{lcccccccccc}
        \toprule
        \multicolumn{4}{c}{GLS} &
        \multicolumn{4}{c}{GLSp} &
        \multicolumn{3}{c}{WWZ} \\
        \cmidrule(lr){1-4} \cmidrule(lr){5-8} \cmidrule(lr){9-11}
        N & Frequency & Period & Ident. &
        N & Frequency & Period & Ident. &
        N & Frequency & Period \\
        & [d$^{-1}$] & [d] & & & [d$^{-1}$] & [d] & & & [d$^{-1}$] & [d] \\
        \midrule
        $F_1$ & 0.0551 [1] & $\mathbf{18.2} \pm 2.8$ & -- &
        $F_1$ & 0.0381 [5] & $26.2 \pm 3.0$ & -- &
        $F_1$ & 0.0537 [1] & $\mathbf{18.6} \pm 2.8$ \\

        $F_2$ & 0.0856 [3] & $\mathbf{11.7} \pm 1.7$ & -- &
        $F_2$ & 0.0541 [1] & $\mathbf{18.5} \pm 2.8$ & -- &
        $F_2$ & 0.0877 [3] & $\mathbf{11.4} \pm 1.0$  \\

        $F_3$ & 0.1294 [4] & $\mathbf{7.7} \pm 0.5$ & -- &
        $F_3$ & 0.0743 [6] & $13.5 \pm 0.9$ & $2 \times F_1$ &
        $F_3$ & 0.1290 [4] & $\mathbf{7.8} \pm 0.5$  \\

        $F_4$ & 0.1862 [2] & $5.4 \pm 0.2$ & $F_1 + F_3$ &
        $F_4$ & 0.0915 [4] & $\mathbf{10.9} \pm 0.9$ & -- &
        $F_4$ & 0.1802 [2] & $5.6 \pm 0.9$  \\

        $F_5$ & 0.2305 [5] & $4.3 \pm 0.3$ & $4 \times F_1$ &
        $F_5$ & 0.1258 [3] & $\mathbf{8.0} \pm 0.5$ & $F_2 + F_3$ &
        $F_5$ & 0.2298 [5] & $4.4 \pm 0.3$  \\

        $F_6$ & 0.5263 [6] & $1.9 \pm 0.1$ & $4 \times F_3$ &
        $F_6$ & 0.1862 [2] & $5.4 \pm 0.9$ & $2 \times F_4$ &
        --    & --     & --            \\

        $F_7$ & 0.6250 [7] & $1.6 \pm 0.1$ & -- &
        $F_7$ & 0.2841 [7] & $3.5 \pm 0.1$ & -- &
        --    & --     & --            \\
        \bottomrule
    \end{tabular}
    \tablefoot{Results of the analysis using three different methods, GLS, GLSp, and WWZ, arranged by decreasing period in each column. The number in brackets [N] denotes the rank of the period by detected power.}
\end{table*}

\begin{figure}
    \centering
    \includegraphics[width=\linewidth]{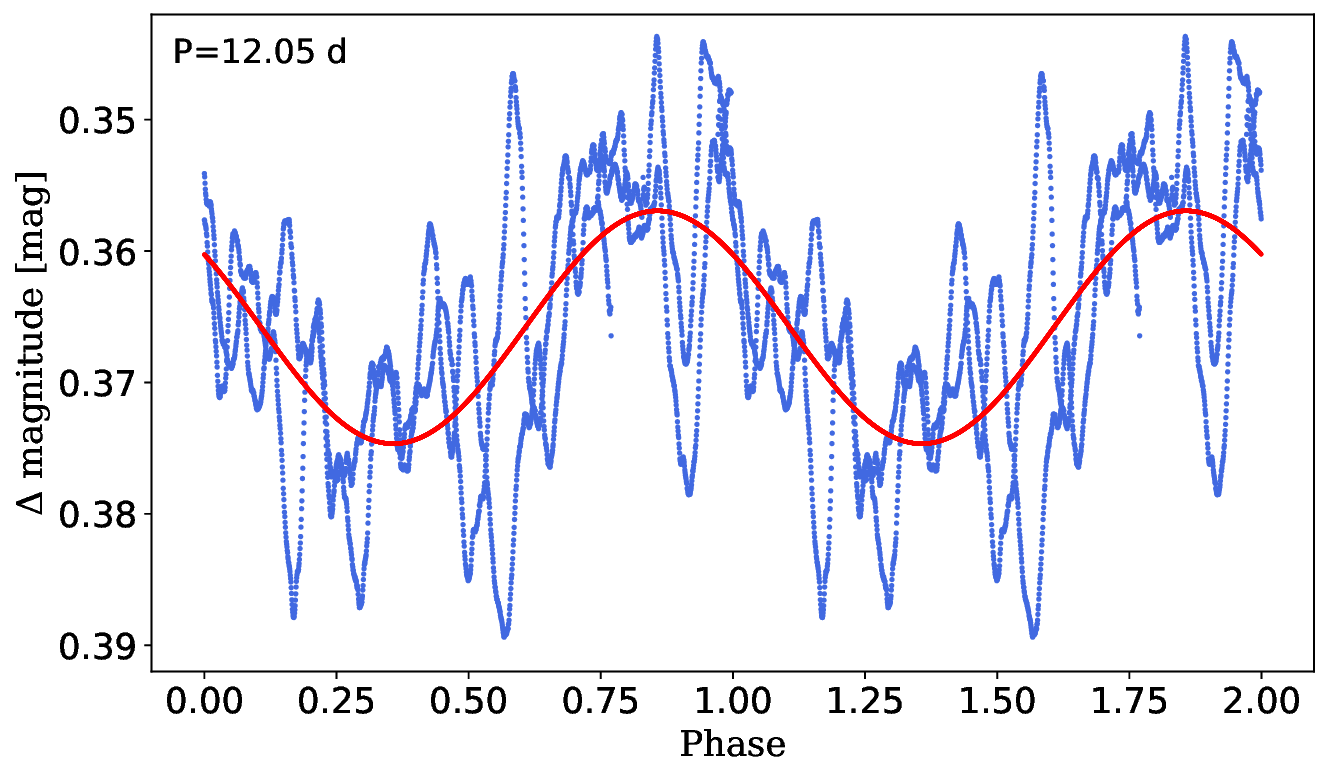}
    \caption{One of the most powerful periods from GLS pre-whitened analysis for the TESS sector 46. Pre-whitened data 
    phased to the indicated period.} 
    \label{fig:TESS 46} 
\end{figure}

\begin{figure*}
    \centering
    \includegraphics[width=\linewidth]{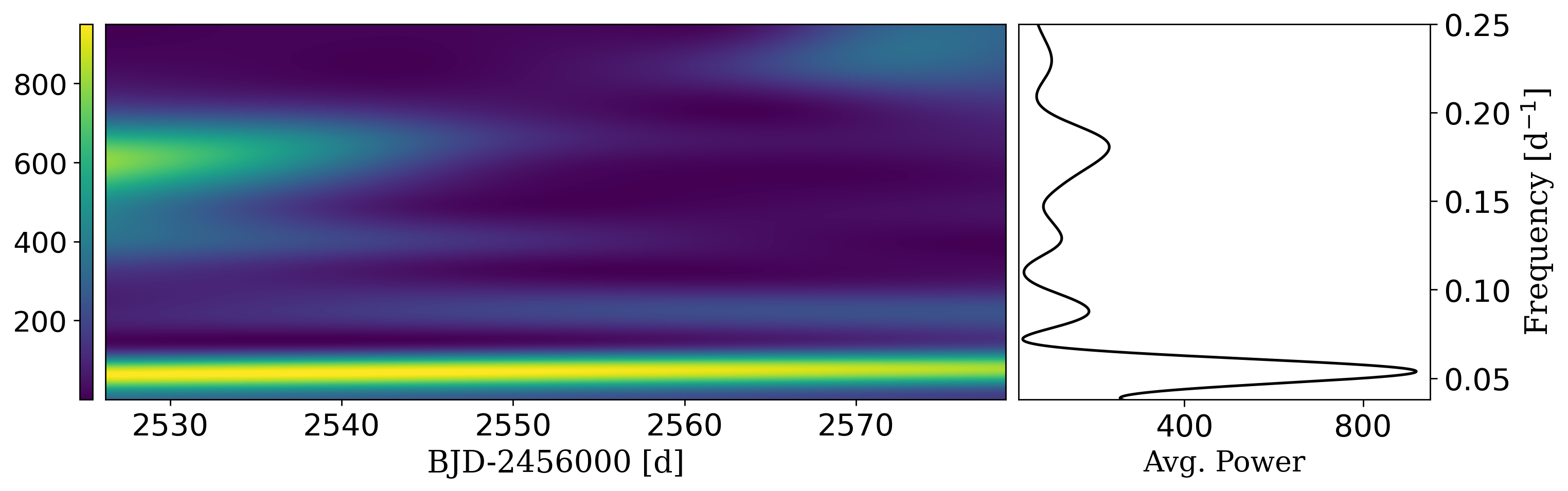}
    \caption{WWZ scalogram for TESS photometry. The wavelet power is indicated by the colour (see colour bar at left). The lower bright stripe corresponds to the period of $\sim$18 days.  A faint stripe appears above it, which corresponds to a period of $\sim$11 days. Shorter periods of  8 and  5.4 days appear as non-permanent spot-like regions in the upper part on the colour scalogram. The black solid line in the right panel shows the time-averaged wavelet power.} 
    \label{fig:WWZ TESS} 
\end{figure*}

By constructing a phase diagram, we can illustrate how the period aligns with our data. Moreover, it enables us to pinpoint a more accurate value for the period. Hence, the next step is GLSp. In Table~\ref{tab:TESS_combined_sorted} we present the set of frequencies that we found. The dominant frequency is approximately 0.054 d$^{-1}$, corresponding to a period of $P \sim 18.5$ days. 
The frequency $F_4 \sim 0.092$ d$^{-1}$ ($P_4 \sim 11$ days) is close to the GLS-identified $F_2$. The analysis was carried out using data from sectors 45 and 46 combined. When analysed separately, sector 46 data give a slightly longer period of $12.1\pm1.3$ days, consistent within the uncertainties. For clarity of presentation, we show the phase diagram for sector 46 using the period specific to that sector (Fig. \ref{fig:TESS 46}), where the signal appears more clearly.
The remaining signals can be interpreted as sub-harmonics or combinations of independent frequencies (e.g. $F_3 \sim 2 \times F_1$ and $F_5 \sim F_2 + F_3$). A weak signal at the frequency corresponding to the period of $26\pm 3$ days, obtained using the GLSp method, is close to the dominant frequency reported by \citet{10.1093/mnras/sty308} based on K2 photometry, and also identified in our analysis of the same dataset.

Finally, we present the WWZ scalogram to understand the temporal evolution of the frequencies and find quasi-periodic events (Figure \ref{fig:WWZ TESS}). We found that the most significant frequencies are in the low-frequency range. Therefore, we provide here the resulting diagram with a range from 0.038 to 0.25, which corresponds to the period range from 4 to 26.22 days. The upper limit of 26.2 days is linked to the time series length of TESS photometry (sectors 45 and 46 in total had an observation period of 52.44 days). 

The strongest signal, with a period of 18.6 days, is relatively stable. Its power slowly decreases towards the end of the observing sequence, while the frequency shows a slight increase. Although less sharply defined, the 11.4-day period remains consistently present throughout. The 7.8-day period is clearly detected at the beginning of the observations, but its power declines rapidly, becoming almost undetectable by the end of the sequence. Additional shorter periods appear intermittently, mainly at the beginning or end of the observing sequence.

\subsubsection{Analysis of the spectroscopic data}\label{S-specana}

Our photometric analyses are complemented with the TO spectra obtained close in time with TESS and K2 data, providing the possibility to compare the results. As mentioned in Section~\ref{S-specobs}, we focus primarily on two observational seasons: 2017 and 2022. 

The spectroscopic frequency analysis is based on the \ion{He}{I}\, 6678.151\,\AA\ line (hereafter the \ion{He}{I} line). We examined the temporal variations in the profile of this line and performed an analysis of its first moment $\langle v_{\rm1} \rangle$ (called the centroid), which represents changes in the radial velocity of a star, in accordance with   Eq. 5.67 in \citet{2010aste.book.....A}.
To investigate the frequency content of the spectral line variability, we adopted an average radial velocity of the system's centre of mass, \vr\ = 42 \kms, as the zero point on the wavelength scale. 
The same methods as for the light curve analyses, GLS, GLSp, and WWZ, were exploited. 

\letteredsubsection{Season 2017}\label{S-specana2017}
In Fig.~\ref{M1_2017}, we present the first-moment measurements for Season 2017 from our 369 spectra collected during 20 nights in the period 4~January~2017 to 16~May~2017. In that season we did not have large gaps between observational runs in the second half of the observing period. Therefore, we can take quite a long frequency range for analysis. The lower limit of 0.0076 d$^{-1}$ corresponds to $1/T$, where T is the observing period (131.8 days). The upper frequency limit was taken as 1.2 d$^{-1}$ because no high frequencies with a sufficient signal-to-noise ratio were found.

\begin{figure}[ht]
    \centering
    \includegraphics[width=\linewidth]{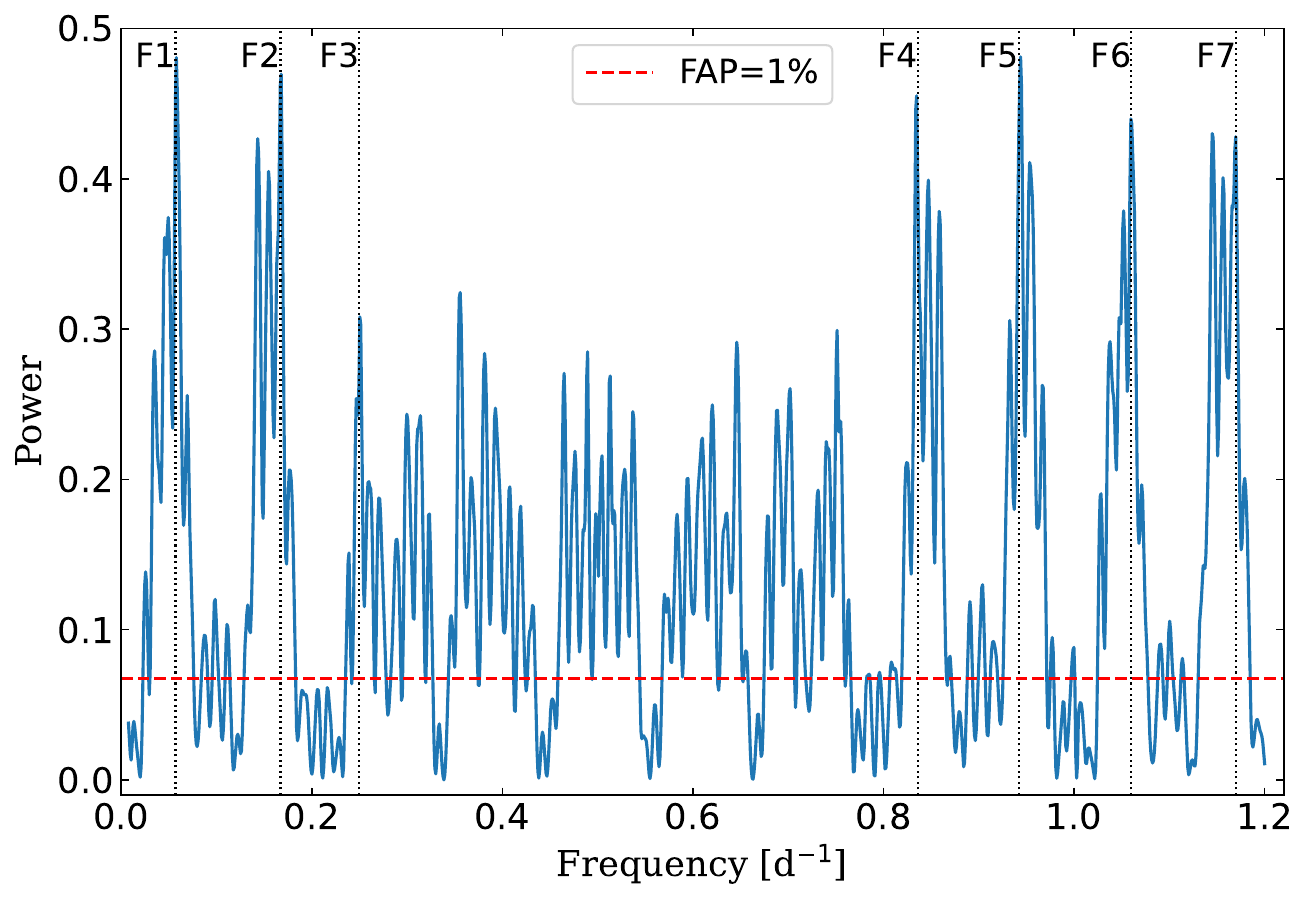}
    \caption{GLS power spectrum for  Season 2017.} 
    \label{fig:LoSc 2017}
\end{figure}

First, the GLS method was applied. The power spectrum is presented in Fig.~\ref{fig:LoSc 2017} and the corresponding periods in Table~\ref{tab:2017_combined}. We note that the general view of the periodogram differs from the same analysis of photometric data. In addition,  several periods close to 1 day can be seen. Most probably, these periods are connected with observational selection because we had a number of observing nights in a row.

\setlength{\tabcolsep}{5pt}
\begin{table*}[h!]
    \captionsetup{justification=raggedright,singlelinecheck=false}
    \centering
    \caption{Periods detected in  Season 2017.}
    \label{tab:2017_combined}
    \begin{tabular}{lcccccccccc}
        \toprule
        \multicolumn{4}{c}{GLS} &
        \multicolumn{4}{c}{GLSp} &
        \multicolumn{3}{c}{WWZ} \\
        \cmidrule(lr){1-4} \cmidrule(lr){5-8} \cmidrule(lr){9-11}
        N & Frequency & Period & Ident. &
        N & Frequency & Period & Ident. &
        N & Frequency & Period \\
        & [d$^{-1}$] & [d] & & & [d$^{-1}$] & [d] & & & [d$^{-1}$] & [d] \\
        \midrule
        
        $F_1$ & 0.0578 [2] & $\mathbf{17.3} \pm 2.1$ & -- &
        $F_1$ & 0.0581 [1] & $\mathbf{17.21} \pm 1.19$ & -- &
        $F_1$ & 0.0344 [4] & $29.1 \pm 7.0$ \\
        
        $F_2$ & 0.1672 [3] & $6.0 \pm 0.3$ & -- &
        $F_2$ & 0.0866 [6] & $\mathbf{11.55} \pm 0.96$ & -- &
        $F_2$ & 0.0455 [3] & $22.0 \pm 2.8$ \\
        
        $F_3$ & 0.2500 [7] & $4.0 \pm 0.2$ & -- &
        $F_3$ & 0.0911 [4] & $\mathbf{10.98} \pm 0.95$ & -- &        
        $F_3$ & 0.0575 [1] & $\mathbf{17.4} \pm 2.7$ \\

        $F_4$ & 0.8360 [4] & $1.2 \pm 0.1$ & -- &
        $F_4$ & 0.1667 [2] & $6.00 \pm 0.22$ & $F_1+F_3$ &
        $F_4$ & 0.0714 [6] & $14.0 \pm 1.5$ \\
        
        $F_5$ & 0.9424 [1] & $1.1 \pm 0.0$ & -- &
        $F_5$ & 0.3157 [5] & $3.20 \pm 0.70$ & $2 \times F_4$ &
        $F_5$ & 0.0855 [7] & $\mathbf{11.7} \pm 0.8$ \\
        
        $F_6$ & 1.0640 [5] & $0.9 \pm 0.0$ & -- &
        $F_6$ & 0.7828 [3] & $1.28 \pm 0.01$ & -- &
        $F_6$ & 0.1429 [2] & $\mathbf{7.0} \pm 0.3$ \\

        $F_7$ & 1.1704 [6] & $0.8 \pm 0.0$ & -- &
        $F_7$ & 1.1765 [7] & $0.85 \pm 0.01$ & -- &
        $F_7$ & 0.1667 [5] & $6.0 \pm 0.3$ \\
        \bottomrule
    \end{tabular}
    \tablefoot{Results of the analysis using three different methods, GLS, GLSp, and WWZ, arranged by decreasing period in each column. The number in brackets [N] denotes the rank of the period by detected power.}
\end{table*}

\begin{figure*}[ht]
    \centering
    \begin{minipage}[t]{0.48\textwidth}
        \centering
        \includegraphics[width=\linewidth,height=0.35\textheight,keepaspectratio]{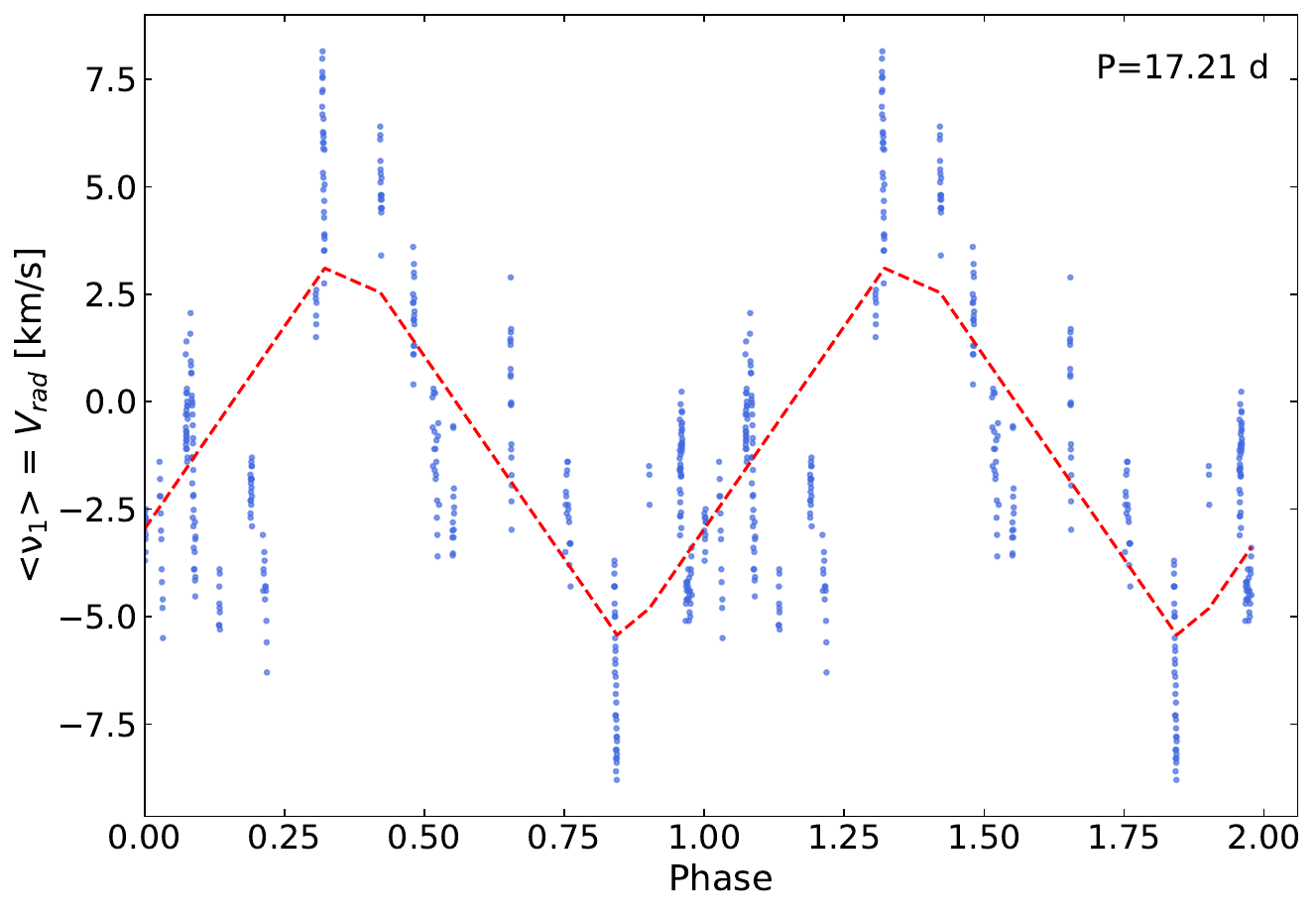}
    \end{minipage}
    \begin{minipage}[t]{0.492\textwidth}
        \centering
        \includegraphics[width=\linewidth,height=0.35\textheight,keepaspectratio]{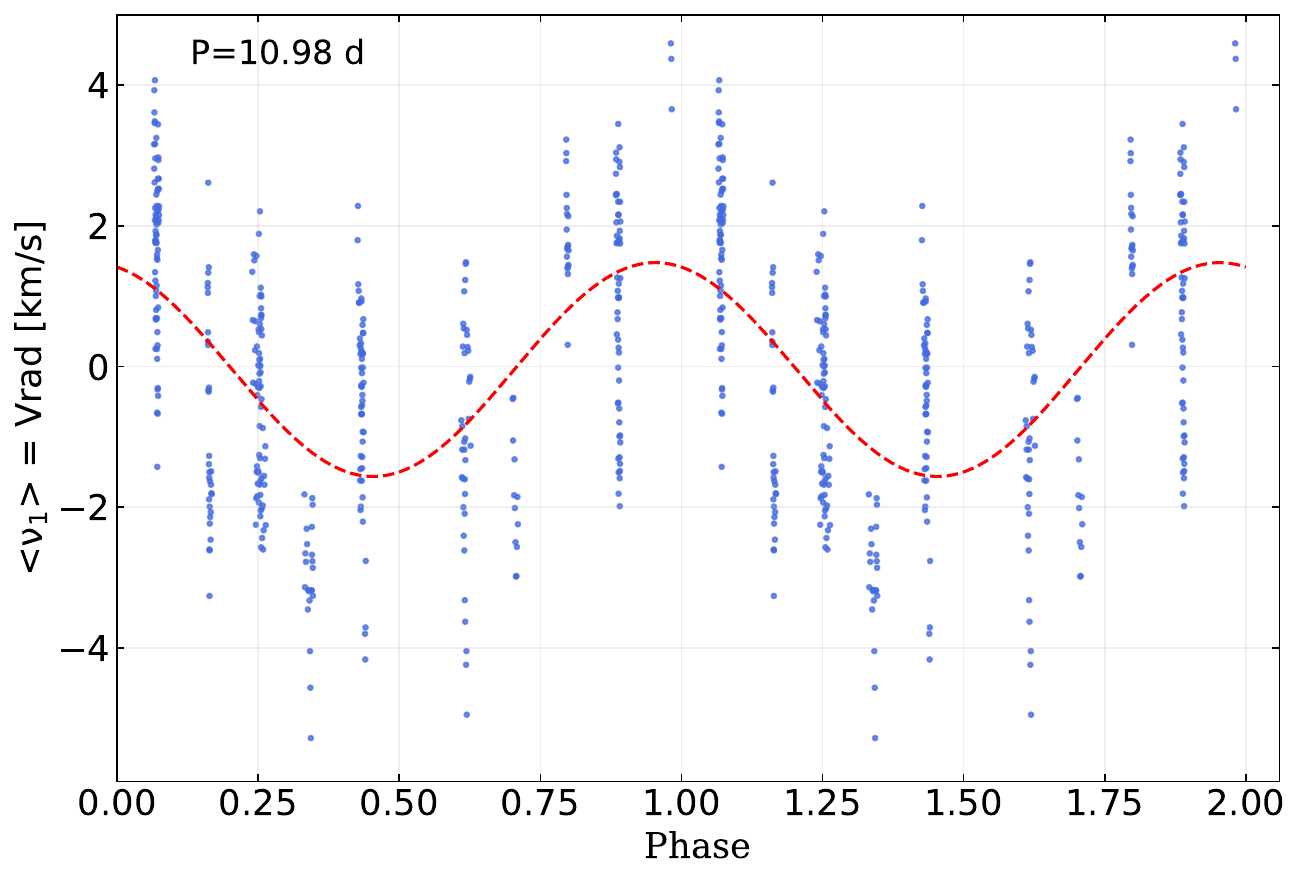}
    \end{minipage}
    \caption{Phase diagrams of  Season 2017 obtained with the GLSp method.}
    \label{fig:2017-GLSp}
\end{figure*}

\begin{figure*}[ht]
    \centering
    \includegraphics[width=\linewidth]{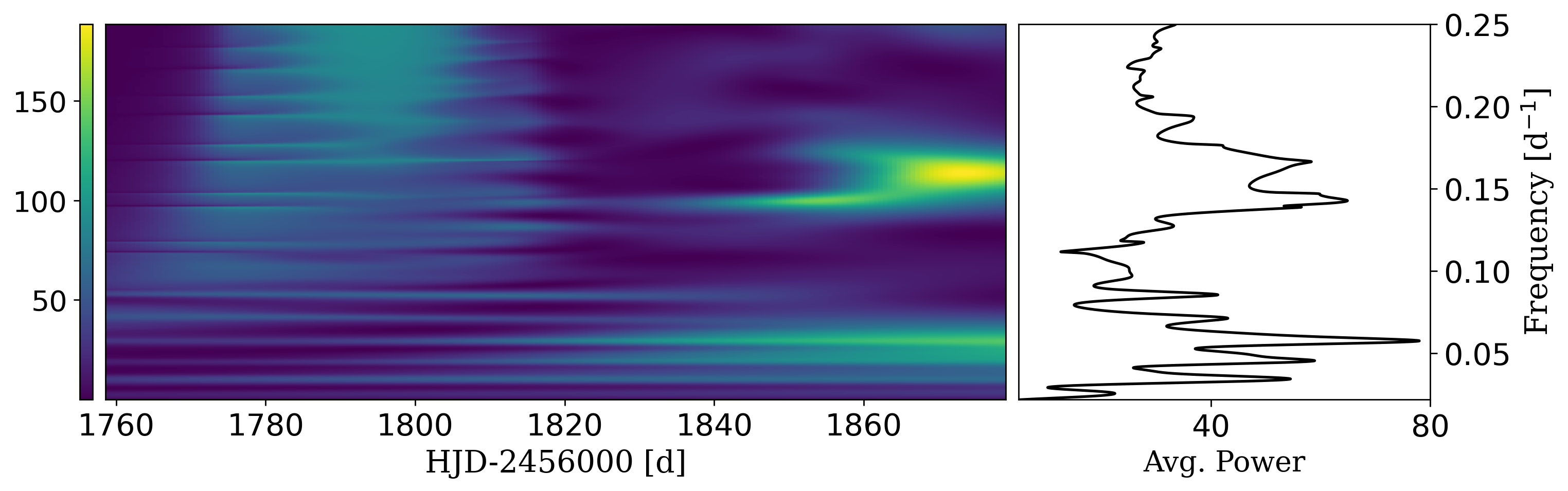}
    \caption{Result of the WWZ analysis for Season 2017. The black solid line in the right panel shows the time-averaged wavelet power.} 
    \label{fig:WWZ M1 2017}
\end{figure*}

Secondly, to derive the periods and their powers, we performed the GLSp. The periods found are listed in Table~\ref{tab:2017_combined}. We note that the fourth frequency $F_{\mathrm{4}}$ (0.1667) is roughly equal to the sum of $F_{\mathrm{1}}$ (0.0582) and $F_{\mathrm{3}}$ (0.0911), also \hbox{$F_{\mathrm{5}}$ (0.3157) $\sim$ the twofold frequency $F_{\mathrm{4}}$}. Phase-folded  plots of the two persistent periods ($\sim$ 17 and 11 days) are shown in Fig.~\ref{fig:2017-GLSp}. We note that the folded velocity curve with a period of 17.21 days is not sinusoidal but has a sawtooth-like shape, whereas the curve with a period of 10.98 days can be well approximated by a sine wave. 

It is difficult to determine the best-fit curve by eye. To quantify the asymmetry of the phase curve, we compared two simple models: a pure sinusoid and a piecewise linear (sawtooth) function with the same period. Both models were fitted to the phase-folded velocity data using least-squares minimization. The goodness of each fit was evaluated using the root mean square (RMS) of the residuals, as well as the Akaike information criterion (AIC) \citet{1974ITAC...19..716A} and the Bayesian information criterion (BIC) \citet{10.1214/aos/1176344136}, consistent with a small but measurable asymmetry between the rising and declining branches. The differences for the two fits for the phase curve with a period of 17.21 days are $\Delta \text{AIC}\sim48$ and $\Delta\text{RMS}\sim 0.16$, which means that sawtooth fitting better describes the behaviour of the phase curve.

Lastly, we performed the WWZ analysis to understand the behaviour of the frequencies and find the quasi-periodic events (Fig.\ref{fig:WWZ M1 2017}). The range of periods was taken from 4 to 61.4 days because this range appeared to be the most informative for this type of analysis. In addition to several long periods in the lower part of the plot, one short period is also visible in the form of a fuzzy spot in the upper left part. The results obtained from the time-averaged power spectrum are given in Table~\ref{tab:2017_combined}. As this method gives only rough frequency estimates, the uncertainties are much higher than for the GLSp method. It is clearly visible on the scalogram that two periods, $\sim6$ days and $\sim7$ days that presented as a yellow spot on the right side near 1870 HJD, do not exist all the time, but only at the end of the season. Several long periods are at the bottom, and the strongest of them is 17.4 days. The period of $\sim11.7$ days, already identified in other analyses, is also present in this dataset. We note that while most of the results obtained by the WWZ analysis agree well with the GLS results (see Table~\ref{tab:2017_combined}), the two longest periods, 22 and $\sim29$ days, are not detected by any of the GLS analyses. These two periods on the scalogram have an unclear shape and are not constant in time. This suggests that they are likely false. Consequently, we do not consider them further in our analysis.

\letteredsubsection{Season 2022}\label{S-specana2022}
Spectroscopic observations in Season 2022 were made mostly right after the TESS photometry. However, one observing night was just before TESS, 1~November~2021, and two during the  TESS observations, 21~November~2021 and 09 December 2021. In Fig.~\ref{M1_2022}, we present the first-moment measurements for that season. As shown in the figure, we obtained a large number of spectra during  Season 2022 (1106 over 39 nights), which enabled us to perform a frequency analysis over a broad range with high confidence. The lower limit 0.005 d$^{-1}$ corresponds to $1/T$, where T is the observing period (194.7 days), and the upper limit is 1.2~d$^{-1}$.

\begin{figure*}[htbp]
    \centering
    \begin{subfigure}[t]{0.495\linewidth}
        \centering
        \includegraphics[width=\linewidth,height=0.35\textheight,keepaspectratio]{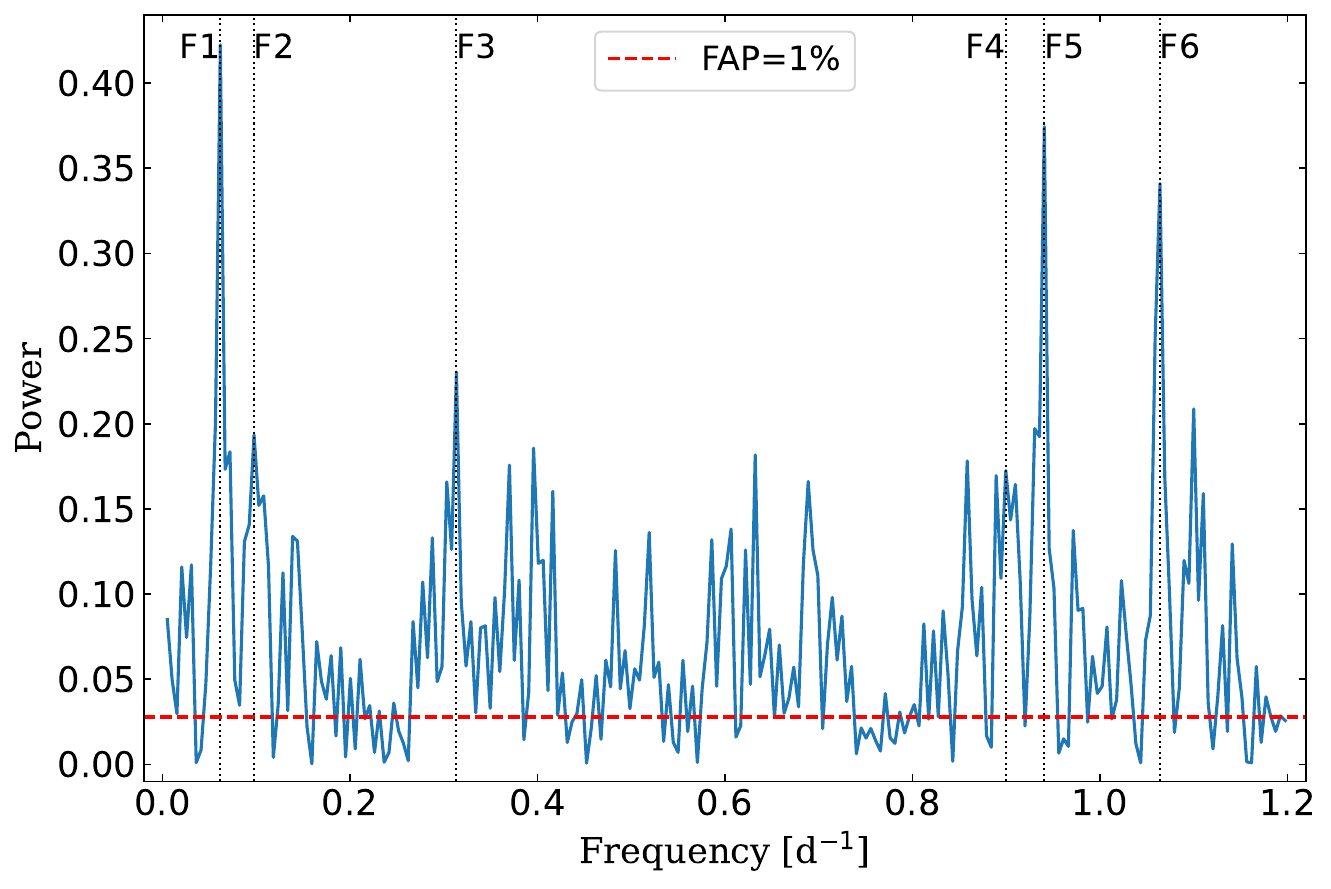}
        \caption{}
        \label{fig:2022-GLS}
    \end{subfigure}
    \hfill
    \begin{subfigure}[t]{0.47\linewidth}
        \centering
        \includegraphics[width=\linewidth,height=0.35\textheight,keepaspectratio]{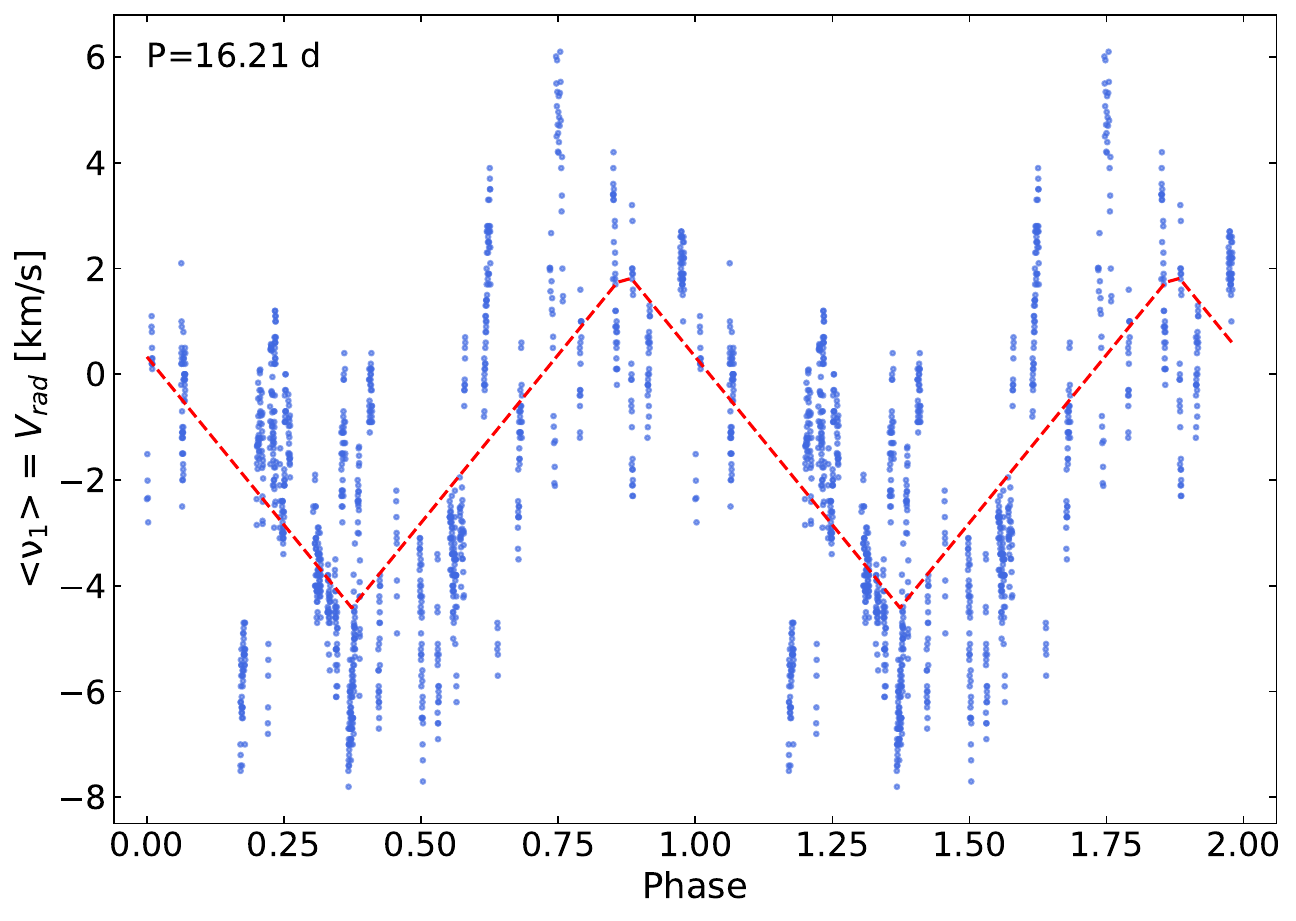}
        \caption{}
        \label{fig:Phase2022}
    \end{subfigure}
    \caption{Analysis of spectroscopic variability in 2022: (a) GLS periodogram; (b) Phase diagram with the most powerful period.}
    \label{fig:2022_combined}
\end{figure*}

As with other analyses, we started with the simple GLS periodogram without pre-whitening. The corresponding power spectrum is presented in Fig.~\ref{fig:2022-GLS}. We determined that all the significant frequencies lie below 1.2 d$^{-1}$. As in the previous analysis, we determined the first seven frequencies (see Table~\ref{tab:2022_combined}). A period of $\sim10$ days is also observed, taking into account the uncertainties in this season. The most powerful period of the spectroscopic data is $\sim16.2$ days. It should be noted that a general view of the periodogram is different from the power spectrum of the TESS photometry (Fig.~\ref{fig:TESS_GLS}).

\setlength{\tabcolsep}{5pt}
\begin{table*}[h!]
    \captionsetup{justification=raggedright,singlelinecheck=false}
    \centering
    \caption{Periods detected in Season 2022.}
    \label{tab:2022_combined}
    \begin{tabular}{lcccccccccc}
        \toprule
        \multicolumn{4}{c}{GLS} &
        \multicolumn{4}{c}{GLSp} &
        \multicolumn{3}{c}{WWZ} \\
        \cmidrule(lr){1-4} \cmidrule(lr){5-8} \cmidrule(lr){9-11}
        N & Frequency & Period & Ident. &
        N & Frequency & Period & Ident. &
        N & Frequency & Period \\
        & [d$^{-1}$] & [d] & & & [d$^{-1}$] & [d] & & & [d$^{-1}$] & [d] \\
        \midrule
        $F_1$ & 0.0617 [1]  & $\mathbf{16.2} \pm 1.9$ & -- &
        $F_1$ & 0.0617 [1]  & $\mathbf{16.21} \pm 1.85$ & -- &
        $F_1$ & 0.0292 [3] & $34.3 \pm 7.0$ \\
        
        $F_2$ & 0.0976 [6] & $\mathbf{10.2} \pm 0.6$ & -- &
        $F_2$ & 0.3033 [7] & $3.30 \pm 0.01$ & -- &
        $F_2$ & 0.0610 [1] & $\mathbf{16.4} \pm 1.8$ \\

        $F_3$ & 0.3135 [4] & $3.2 \pm 0.0$ & $3 \times F_2$ &
        $F_3$ & 0.5294 [6] & $1.89 \pm 0.07$ & -- &
        $F_3$ & 0.1000 [2] & $\mathbf{10.0} \pm 1.1$ \\
        
        $F_4$ & 0.8995 [7] & $1.1 \pm 0.1$ & -- &
        $F_4$ & 0.5859 [3] & $1.71 \pm 0.01$ & -- &
        $F_4$ & 0.1389 [4] & $\mathbf{7.2} \pm 0.5$ \\
        
        $F_5$ & 0.9406 [2] & $1.1 \pm 0.0$ & $3 \times F_3$ &
        $F_5$ & 0.8327 [2] & $1.2 \pm 0.07$ & $F_2 + F_3$ &
        $F_5$ & 0.1818 [5] & $5.5 \pm 0.2$ \\

        $F_6$ & 1.0639 [3] & $0.9 \pm 0.0$ & -- &
        $F_6$ & 0.8584 [5] & $1.17 \pm 0.07$ & -- &
        $F_6$ & 0.2128 [3] & $4.7 \pm 0.2$ \\

        $F_7$ & -- & -- & -- &
        $F_7$ & 1.1205 [4] & $0.89 \pm 0.01$ & -- &
        --    & -- & -- \\

        \bottomrule
    \end{tabular}
    \tablefoot{Results of the analysis using three different methods, GLS, GLSp, and WWZ, arranged by decreasing period in each column. The number in brackets [N] denotes the rank of the period by detected power.}
\end{table*}

To derive the periods and build the phase-folded diagrams, we performed the GLSp;   the resulting periods are listed in Table~\ref{tab:2022_combined}. We note that $F_{\mathrm{5}}$ $\sim$ $F_{\mathrm{2}}$ + $F_{\mathrm{3}}$.
In Fig.~\ref{fig:Phase2022}, we present a phase plot with the first most powerful period we  found. The curve exhibits a pronounced sawtooth shape. 

\begin{figure*}
    \centering
    \includegraphics[width=\linewidth]{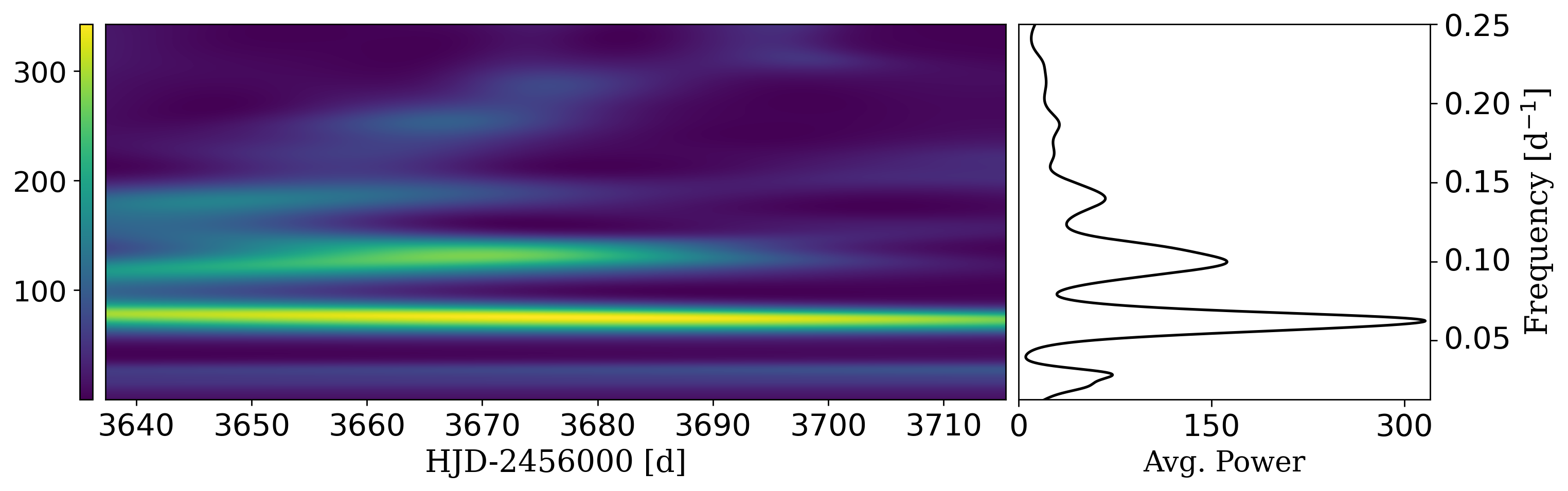}
    \caption{Result of the WWZ analysis for Season 2022. The wavelet power is indicated by the colour (see colour bar at left). The black solid line in the right panel shows the time-averaged wavelet power.} 
    \label{fig:WWZ M1 2022}
\end{figure*}

Lastly, we obtained a very interesting result using the WWZ analysis (Fig. \ref{fig:WWZ M1 2022} and Table~\ref{tab:2022_combined}). The frequency range was taken from 0.01 to 0.25 d$^{-1}$. The lowest and strongest frequency on the scalogram corresponds to the period $\sim16.4$ days. According to the scalogram, this period was very pronounced and existed throughout the entire observation season and slightly increases by the end of the season. The period of $\sim 11$ days also exists in the scalogram of  Season 2022, and here the value is equal to 10 $\pm$ 1 days, although this period does not exist all the time and increases towards the end, according to the scalogram.

\subsection{Inference of the inclination angle}\label{Zpektr}
\begin{figure}[!h]
    \centering
    \includegraphics[width=\linewidth]{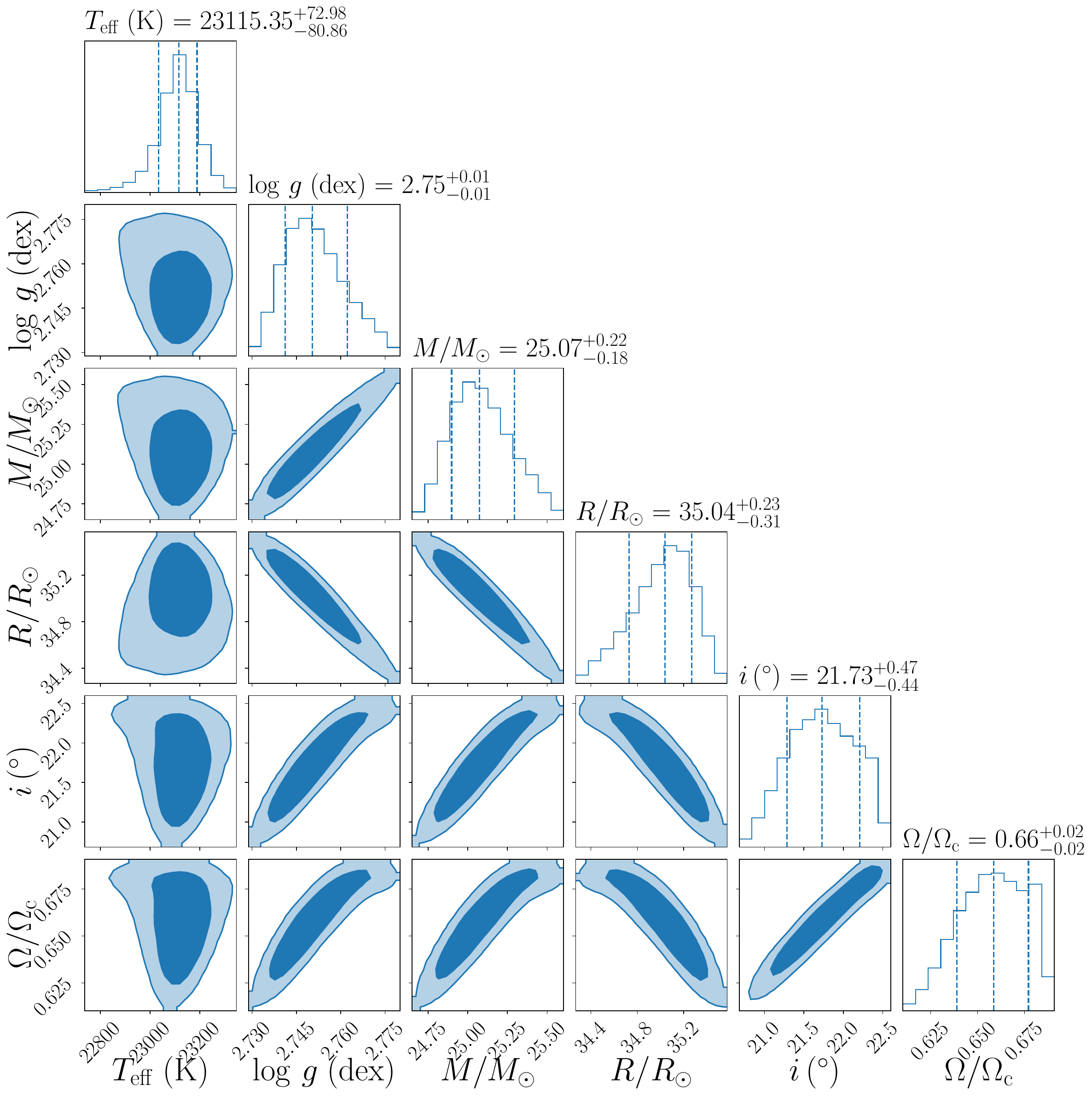}
    \caption{Posterior distributions of the gravity-darkened physical parameters based on \ion{He}{I}\ 4026\,\AA\ line profile modelling with \textsc{zpektr} using Hamiltonian Monte Carlo inference with STAN (HMC). The mean and the 16th and 84th percentile values for each parameter are shown in the histograms, where the blue curves indicate the joint probability density contours for the correlation between the pairs of parameters. The inclination angle that best matches the data on average is ${21.7^{\circ}}^{+0.5}_{-0.4}$} 
    \label{fig:mcmc}
\end{figure}

Assuming the star to be a rigid rotator and that the stellar radiative flux
in the outer atmospheric layers is anti-parallel to the effective gravity to preserve the flux/gravity ratio for all latitudes on the stellar surface, we used the \textsc{zpektr} code \citep{Levenhagen2014ApJ...797...29L,refId0} to build a grid of $\approx$ 40\,000 gravity-darkened model spectra using the Espinosa-Lara gravity-darkening prescription \citep{Espinosa-Lara2011}. The input plane-parallel model spectra to the code \textsc{zpektr} are a hybrid LTE/NLTE set of model spectra that were built with the \textsc{tlusty} and \textsc{synspec} codes \citep{Hubeny1988CoPhC..52..103H,Hubeny1995ApJ...439..875H} in NLTE regime (for models hotter than 15\,000~K) and LTE Kurucz atmosphere models \citep{Kurucz1979} with \textsc{synspec} (for models cooler than 15\,000~K). The helium abundance was set to that of the star (0.2~dex), and the wavelength range was set to match that of the \ion{He}{I}\ 4026 \AA. 

Figure \ref{fig:mcmc} shows the results of the fittings of the observed \ion{He}{I}\ 4026\,\AA\, line profile with the  \textsc{zpektr} models using a Hamiltonian Monte Carlo (HMC) inference with STAN \citep{Stan-JSSv076i01,DUANE-HMC-1987216}. For this, the HARPS spectrum observed on 12~February~2006 was used. The histograms and correlation plots, showing the joint probability density contours, reflect the correlations of each pair of parameters in the simulations. The best angle of inclination obtained in this analysis is ${21.7^{\circ}}^{+0.5}_{-0.4}$.

\section{Discussion}\label{discus}

In this study, we have obtained a wide range of periods, from less than one day   to $\sim$18 days repeating in all datasets (Tables in Section~\ref{S-res} and Table~\ref{tab:Periods_All}). 
The revealed pulsation pattern is more consistent with the theoretical predictions by \cite{2013MNRAS.433.1246S} for post-RSG blue supergiants, suggesting that  \rholeo\ is on the blue loop of its evolution after the RSG stage.

Several periods identified from the TESS and K2 photometry, when considering the associated uncertainties, coincide with those found in the corresponding spectroscopic data. More importantly, a number of periods repeat in all four datasets, as illustrated in Table~\ref{tab:Periods_All}. Among them, the $\sim$17-day period consistently exhibits the strongest power, as shown in the tables in Section~\ref{S-res}. Also notable is the recurring $\sim$11-day period, even though its power is typically weaker. Both of these periods are independent and are not the result of combinations of other periods. We  explore these findings in more detail below.

\begin{table}[h!]
    \captionsetup{justification=raggedright,singlelinecheck=false}
    \centering
    \caption[]{Summary table of detected periods.}
        \label{tab:Periods_All}
        \setlength{\tabcolsep}{5pt}
        \begin{tabular}{c c c c c}
            \hline\hline
            Average [d] & 2017 & K2    & TESS  & 2022  \\
            period&&&& \\
            \hline
            25.5  &  --            &  $25\pm 3.0$ &  $26 \pm 3$            &  --            \\
            16.8  & $16.2\pm1.1$   & $16.3\pm0.2$ & $18.4\pm1.6$   & $16.3\pm1.1$   \\
            10.8  & $11.6\pm0.6$   & $10.0\pm0.8$  & $11.3\pm0.6$  & $10.1\pm0.4$   \\
            7.3   &  $6.9\pm0.2$   & $7.1\pm0.2$   & $7.8\pm0.3$   & $7.2\pm0.1$    \\
            5.6 & $6.0\pm0.1$    & $5.2\pm0.1$   & $5.5\pm0.6$   & $5.5\pm0.2$    \\
            3.4   & $3.2\pm0.7$    & $2.9\pm0.0^\ast$ & $4.3\pm0.2$ & $3.3\pm0.0^\ast$ \\
            1.2   & $0.9\pm0.0\ast$& --            & $1.6\pm0.0^\ast$& $1.1\pm0.0^\ast$ \\
            \hline
        \end{tabular}
        \tablefoot{Table of repeating periods detected in the spectroscopic and photometric datasets. For the spectral periods, the averaged values of three methods (GLS, GLSp, and WWZ) are stated.  An asterisk (*) indicates an uncertainty of less than 0.04 days.}
\end{table}

\begin{figure}[ht]
    \centering
    \includegraphics[width=\linewidth]{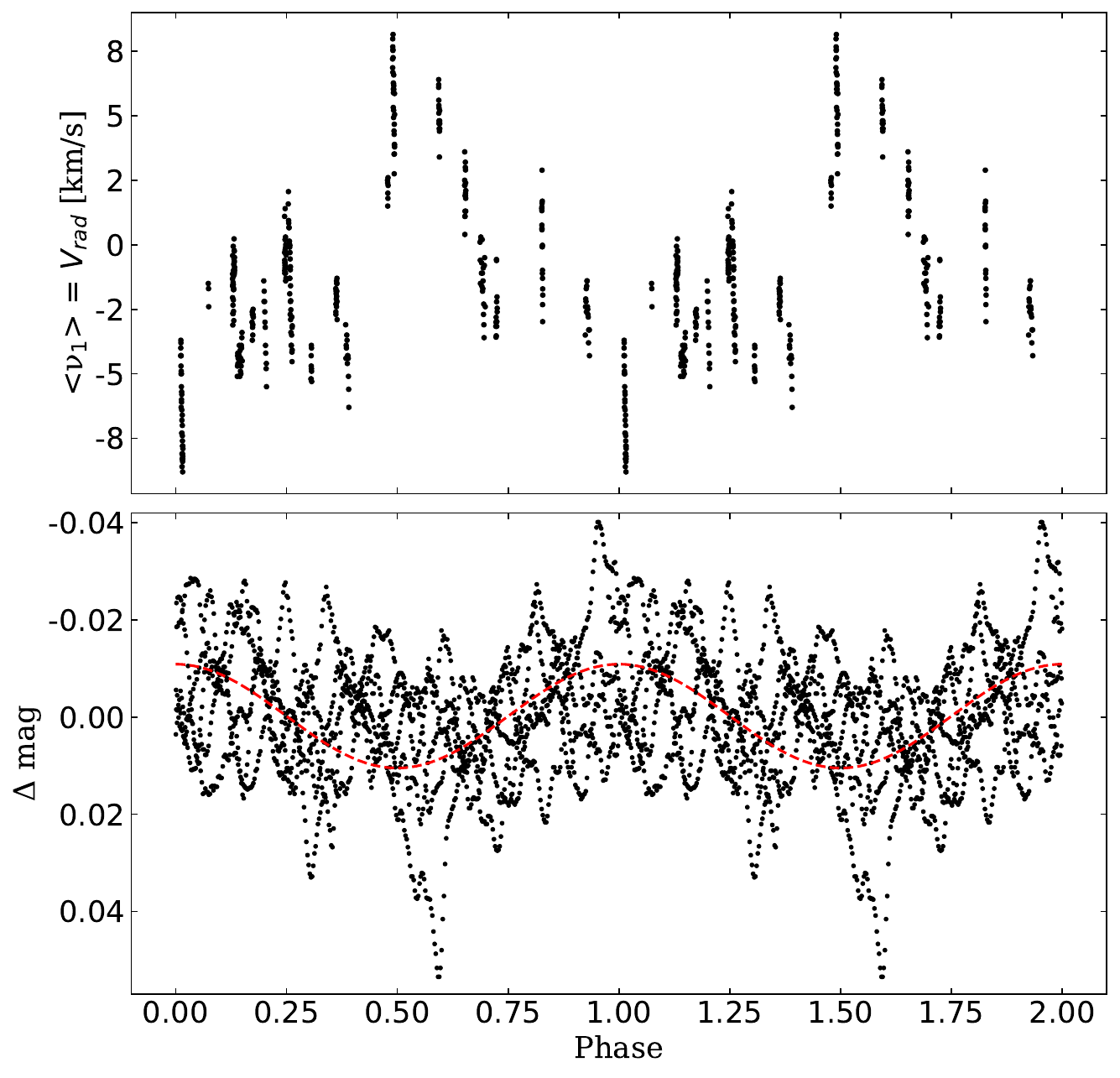}
    \caption{Season 2017 first-moment values (above) and K2 light curve (below) are phase-folded with a period of 17.212 days. The sinusoidal fit to the photometric data is presented in red.}
    \label{fig:2017+K2}
\end{figure}

First, we focused on the period of $\sim17$ days. 
Figure ~\ref{fig:2017+K2} shows the K2 light curve and the first moment $\langle v_{\rm1} \rangle$ of the 2017 spectroscopic data, both folded with a period of 17.212 days. As the observations were taken close in time, we analysed them together. This period gives a visually good fit to both curves, although they are slightly offset from each other. It is also only slightly shorter than the average period derived from the two datasets, 16.29 days.

An interesting feature is the shape of the spectroscopic phase curve;  instead of a sinusoid, it has a sawtooth-like shape.
This could have several possible interpretations. One is that it indicates radial pulsation. Alternatively, the sawtooth shape might suggest the presence of a close binary companion, with periodic mass exchange and possibly a circumbinary disk. These phenomena are consistent with the theoretical models and observational findings reported by \cite{1998AJ....116.1961H} and \cite{2024arXiv240416089P}. Another possibility is that the sawtooth-like shape reflects mass loss modulated by some as-yet-unknown process within the binary system \citep{2008MNRAS.389.1605M}.

Secondly, we turn our attention to the period of \hbox{$\sim11$} days, which  we propose reflects stellar rotation. 
We estimated the inclination angle of the star to be ${21.7^{\circ}}^{+0.5}_{-0.4}$ (Sect.~\ref{Zpektr}). By incorporating the stellar radius and the projected rotational velocity from Table~\ref{tab:Phys.param} into the calculations, we estimate a rotation period of $12.5\pm0.7$ days, which aligns well with the repeating period of $10.8\pm0.3$ days found in all our observational datasets (taking into account the strong dependence of the period on the stellar radius). 

In the literature, two other periods are proposed as being associated with stellar rotation. \cite{2007AN....328.1170K} report a period of 7.3 days. Our analysis of all datasets confirms the existence of this period. 
However, this period is not independent, but is a combination of other frequencies. \citet{10.1093/mnras/sty308} reported a period of 26.8 days based on the combined K2 light curve and high-resolution HERMES spectroscopy, which was mostly obtained well before the K2 observations.
  
We recovered a similar period of approximately 25 days in the K2 data, consistent within the uncertainties with their result, and our WWZ analysis confirms that this signal undergoes amplitude and phase modulation. However, we did not detect it in any of our spectroscopic datasets, and in the TESS photometry it appears only with very low amplitude, below the FAP threshold. This indicates that the reported 26.8-day signal is likely a transient feature rather than the true stellar rotation period.

A periodicity of about 5.5 days, detected across all our datasets, is close to the $f_2$ = 0.1956 d$^{-1}$ ($P$ = 5.11 d)
frequency reported by \citet{10.1093/mnras/sty308} and is likely a harmonic of the 11-day mode. We do not confirm the presence of another frequency, $f_3$ = 0.4588 d$^{-1}$ ($P$ = 2.18 d), which \citet{10.1093/mnras/sty308} attributed to gravity waves. In the K2 data, the corresponding peak lies below the FAP threshold; a similar feature in the TESS photometry is likewise below FAP; and our GLSp analysis of the spectroscopic data from both 2017 and 2022 shows no significant power at this frequency.

\section{Conclusions}\label{conc}

The main outcome of this study is the identification of a set of periods using various methods of frequency analysis. The periods obtained through different techniques are in good agreement and complement each other. Furthermore, there is a clear similarity between the results derived from photometry and spectroscopy. The main findings of our study are summarized as follows:

\begin{enumerate}

      \item We identified a set of repeating (quasi-)periods and their harmonics.
      The recurrence of most of these periods in different seasons supports their authenticity. Several of these periods were detected using multiple methods (GLS, GLSp, and WWZ) and most appear in both photometric and spectroscopic data, further reinforcing their significance.
      
      \item Certain periods appear to be persistent over time, such as those near $\sim$17 days, $\sim$11 days, and $\sim$7 days, suggesting that they may be linked to stable physical processes in the system. We propose that the period of $\sim$17 days is related to radial pulsations. 
          
      \item We provide strong theoretical and observational evidence that the true rotation period of \rholeo\ is $12.5\pm0.7$ days.
            
      \item The wide range of periods obtained may indicate that this star is on the blue loop of evolution, that is, after the RSG stage.

\end{enumerate}

\section{Data availability}\label{data}
Tables \ref{TESS_lc}, \ref{M1_2017}, and \ref{M1_2022} are only available in electronic form at the CDS via anonymous ftp to cdsarc.u-strasbg.fr (130.79.128.5) or via http://cdsweb.u-strasbg.fr/cgi-bin/qcat?J/A+A/

\begin{acknowledgements} 
This work has used the ground-based research infrastructure of Tartu Observatory, funded through the projects TT8 (Estonian Research Council) and KosEST (EU Regional Development Fund). The authors acknowledge support from the Estonian Research Council grant IUT-40.

This project has received funding from the European Union's Horizon Europe research and innovation programme under grant agreement No. 101079231 (EXOHOST), and from the United Kingdom Research and Innovation (UKRI) Horizon Europe Guarantee Scheme (grant number 10051045).

This project has received funding from the European Union’s Framework Programme for Research and Innovation Horizon 2020 under the Marie Skłodowska-Curie Grant Agreement No. 823734 (POEMS) and is also Co-funded by the European Union (Project 101183150 – OCEANS).

AA acknowledges support from the Estonian Research Council grant PRG 2159.

      Part of this work was supported by the German
      \emph{Deut\-sche For\-schungs\-ge\-mein\-schaft, DFG\/} project
      number Ts~17/2--1.

Based on data obtained from the ESO Science Archive Facility with DOI \url{https://doi.eso.org/10.18727/archive/33} under the ESO programme 60.A-9036.

\end{acknowledgements}

\bibliographystyle{aa}

\bibliography{bible}

\begin{appendix}
\twocolumn[
\section{Photometric and spectroscopic time series}\label{timeseries}
]

\begin{figure}
  \centering
  \includegraphics[width=\columnwidth]{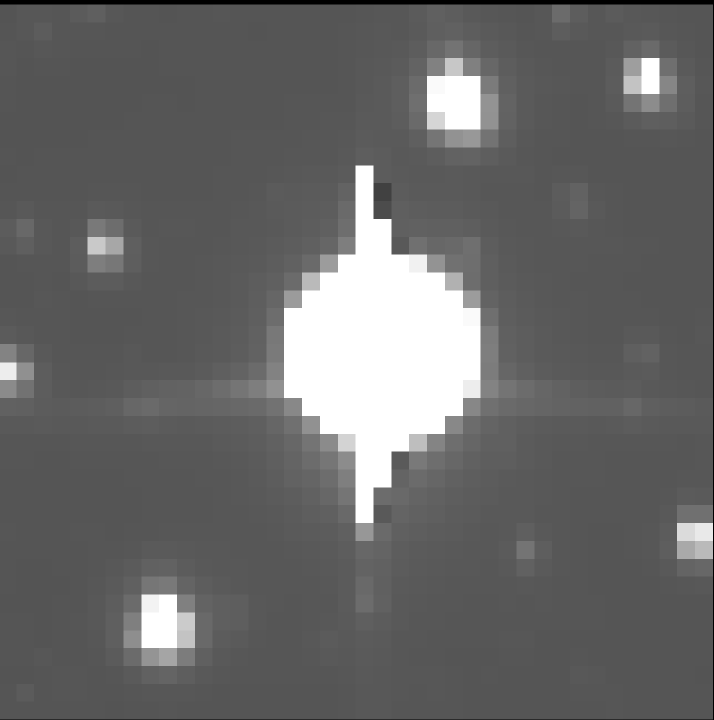}
  \captionof{figure}{Pure-white pixels mark the customized aperture that we use to measure the TESS flux for {\rholeo}, observed from 6 November 2021 to 30 December 2021.}
  \label{TESS_aperture}
\end{figure}

\begin{figure}
  \centering
  \includegraphics[width=\columnwidth]{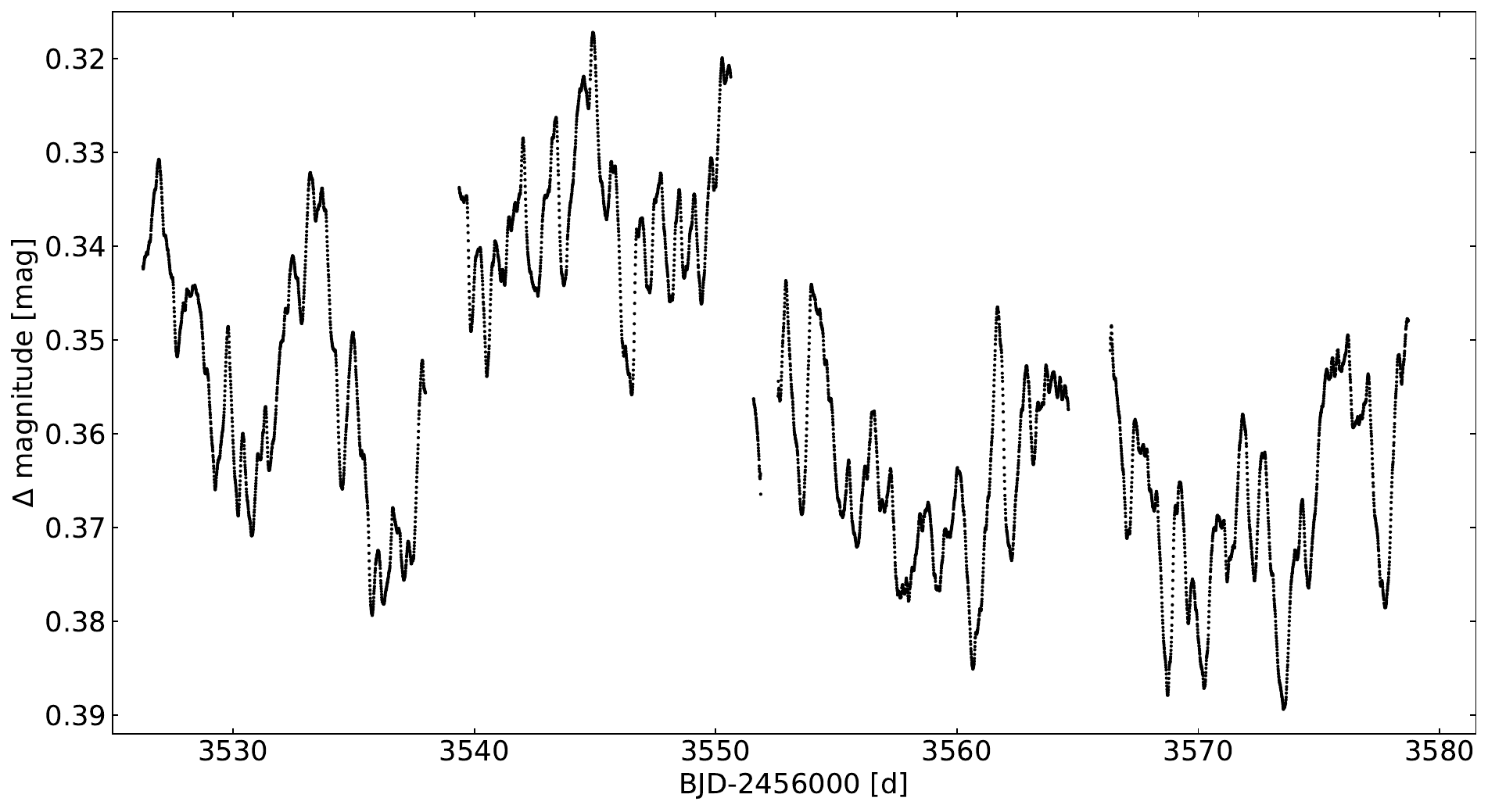}
  \captionof{figure}{TESS photometry of {\rholeo}, observed from 6 November 2021 to 30 December 2021. Short- and long-term variability in brightness are clearly visible. The photometry was obtained just before the 2022 spectroscopic observations. The data are available at CDS.}
  \label{TESS_lc}
\end{figure}

\begin{figure}
  \centering
  \includegraphics[width=\columnwidth]{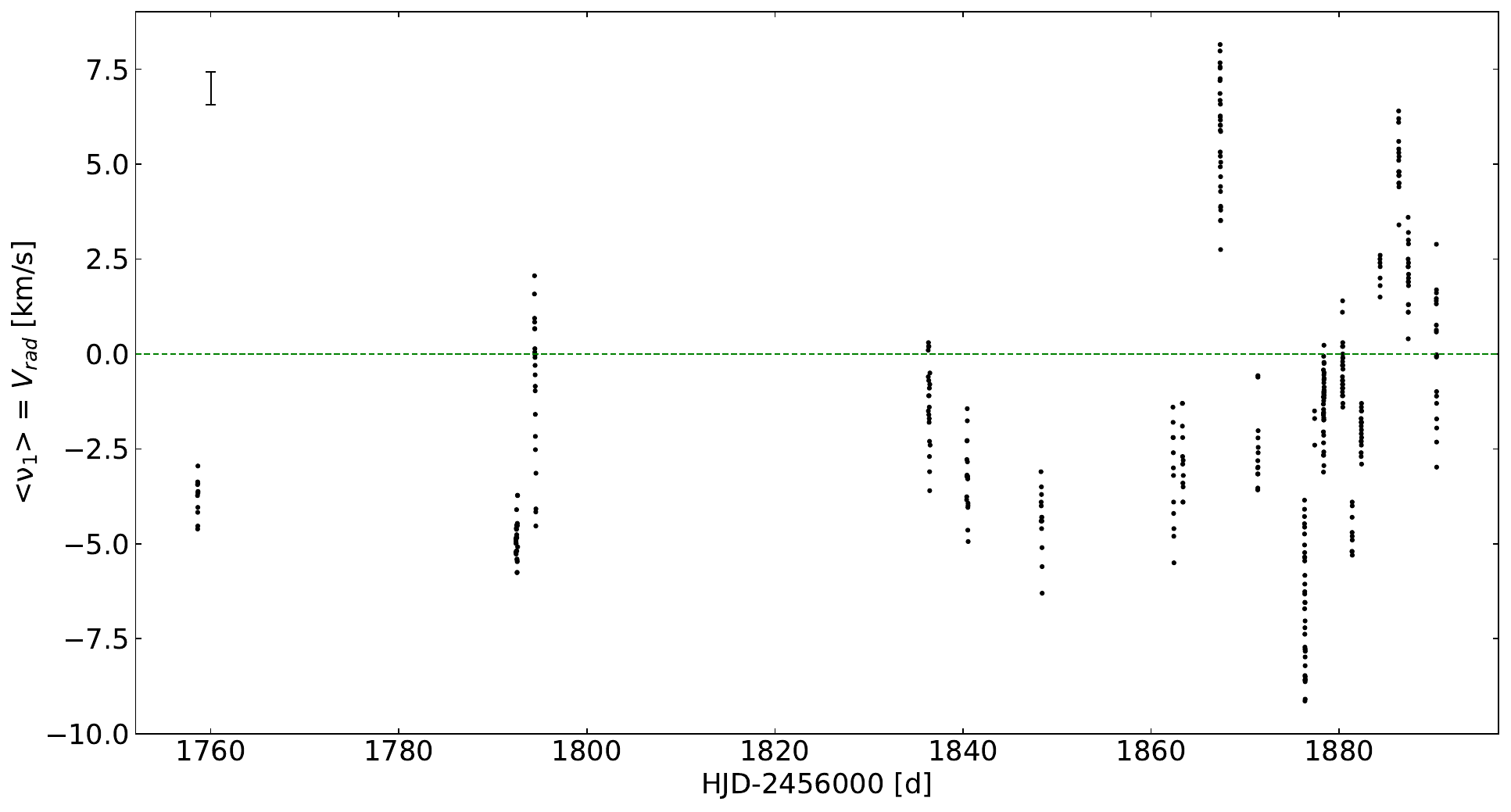}
  \captionof{figure}{First-moment (radial velocity) variations of the \ion{He}{I}\, 6678.151\,\AA\ line profile during  Season 2017 (4~January~2017 -- 16~May~2017). Each point corresponds to one separate spectrum. A typical error bar is shown in the top left corner. The 2017 spectra were obtained just before the K2 campaign. The data are available at CDS.}
  \label{M1_2017}
\end{figure}

\begin{figure}
  \centering
  \includegraphics[width=\columnwidth]{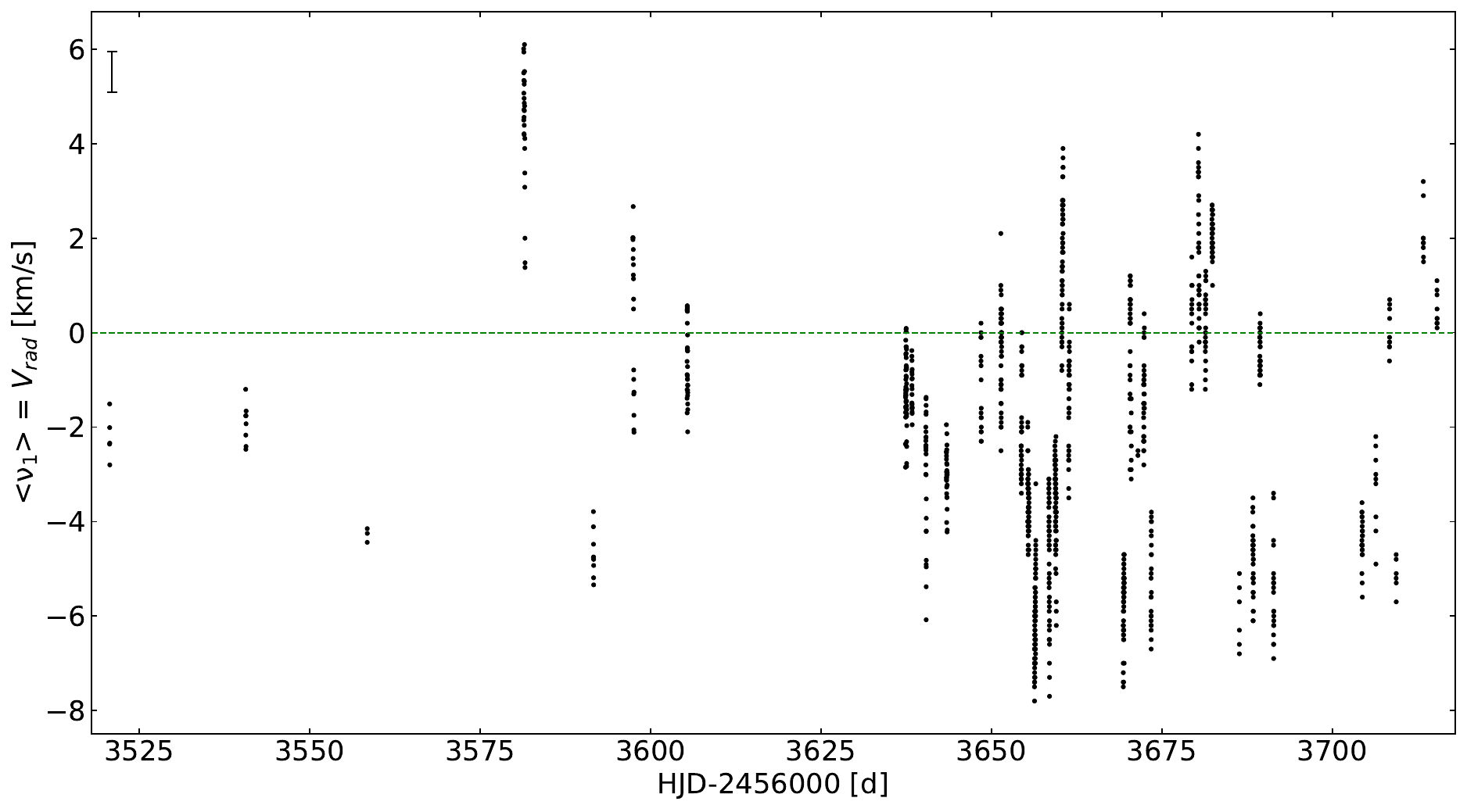}
  \captionof{figure}{Spectroscopic data in the period 1~November~2021 to 15~May~2022 (Season 2022). Each point corresponds to one separate spectrum. In total there are 1106 spectra, 39 nights: one preceding, two coinciding with, and the rest closely following the TESS photometry. The data are available at CDS.}
  \label{M1_2022}
\end{figure}
\clearpage

\onecolumn

\section{Log of spectroscopic observations}\label{S-appendixobs}

\begin{table*}[ht!]
    \captionsetup{justification=raggedright,singlelinecheck=false}
    \centering
    \caption{Spectroscopic data for  Seasons 2017 and 2022.}
    \label{tab:Spectra_all2}
    \begin{tabular}{lrccclrcc}
        \toprule
        \multicolumn{4}{c}{Season 2017} & & \multicolumn{4}{c}{Season 2022} \\
        \cmidrule(lr){1-4} \cmidrule(lr){6-9}
        Night & N & S/N & Exp. [s] & & 
        Night & N & S/N & Exp. [s] \\
        \midrule
        2017-01-04 & 11 & 330 & 350 && 2021-11-01 & 5  & 440 & 640 \\
        2017-02-07 & 25 & 410 & 600 && 2021-11-21 & 8  & 530 & 790 \\
        2017-02-09 & 21 & 390 & 460 && 2021-12-09 & 3  & 310 & 500 \\
        2017-03-23 & 21 & 450 & 640 && 2022-01-01 & 26 & 400 & 620 \\
        2017-03-27 & 18 & 450 & 660 && 2022-01-11 & 9  & 460 & 390 \\
        2017-04-04 & 13 & 520 & 760 && 2022-01-17 & 18 & 500 & 580 \\
        2017-04-18 & 12 & 530 & 640 && 2022-01-25 & 30 & 340 & 240 \\
        2017-04-19 & 12 & 440 & 690 && 2022-02-26 & 50 & 360 & 220 \\
        2017-04-23 & 30 & 320 & 200 && 2022-02-27 & 20 & 390 & 185 \\
        2017-04-27 & 13 & 260 & 250 && 2022-03-01 & 29 & 390 & 165 \\ 
        2017-05-02 & 35 & 220 & 180 && 2022-03-04 & 29 & 390 & 250 \\ 
        2017-05-03 & 10 & 230 & 180 && 2022-03-09 & 17 & 340 & 170 \\ 
        2017-05-04 & 36 & 270 & 160 && 2022-03-12 & 50 & 340 & 150 \\
        2017-05-06 & 24 & 260 & 170 && 2022-03-15 & 40 & 350 & 190 \\
        2017-05-07 & 9  & 330 & 230 && 2022-03-16 & 67 & 270 & 150 \\
        2017-05-08 & 18 & 370 & 210 && 2022-03-17 & 76 & 350 & 160 \\
        2017-05-10 & 7  & 300 & 300 && 2022-03-19 & 42 & 360 & 160 \\
        2017-05-12 & 18 & 350 & 200 && 2022-03-20 & 74 & 350 & 160 \\
        2017-05-13 & 18 & 260 & 190 && 2022-03-21 & 60 & 340 & 180 \\
        2017-05-16 & 19 & 250 & 160 && 2022-03-22 & 36 & 270 & 190 \\
                   &	&     &     && 2022-03-30 & 49 & 330 & 170 \\
                   &	&     &     && 2022-03-31 & 37 & 350 & 150 \\
                   &	&     &     && 2022-04-01 & 2  & 250 & 275 \\
                   &	&     &     && 2022-04-02 & 36 & 350 & 180 \\
                   &	&     &     && 2022-04-03 & 24 & 350 & 160 \\
                   &	&     &     && 2022-04-09 & 16 & 310 & 170 \\
                   &	&     &     && 2022-04-10 & 34 & 360 & 210 \\
                   &	&     &     && 2022-04-11 & 25 & 330 & 160 \\
                   &	&     &     && 2022-04-12 & 38 & 360 & 190 \\
                   &	&     &     && 2022-04-16 & 6  & 330 & 250 \\
   	           &	&     &     && 2022-04-18 & 32 & 340 & 160 \\
   	           &	&     &     && 2022-04-19 & 30 & 330 & 200 \\
   	           &	&     &     && 2022-04-21 & 20 & 340 & 160 \\
	           &	&     &     && 2022-05-04 & 25 & 260 & 200 \\
	           &	&     &     && 2022-05-06 & 9  & 330 & 185 \\
	           &	&     &     && 2022-05-08 & 10 & 360 & 250 \\
	           &	&     &     && 2022-05-09 & 6  & 340 & 165 \\
	           &	&     &     && 2022-05-13 & 9  & 340 & 210 \\
	           &	&     &     && 2022-05-15 & 10 & 270 & 220 \\
\bottomrule
    \end{tabular}
    \tablefoot{1477 spectra in total. The S/N was measured near the \ion{He}{I} line at 6678.151\,\AA{}. S/N and the exposure time values are averaged per night.}
\end{table*}					       

\end{appendix}

\end{document}